\documentclass[12pt]{article}
\usepackage{amssymb}
\usepackage{mathrsfs}
\renewcommand{\theequation}{\arabic{section}.\arabic{equation}}

\begin{document}

%************************** Text Begins here ******************************

%  Greek letters

\def\a{\alpha}
\def\b{\beta}
\def\d{\delta}
\def\e{\epsilon}
\def\g{\gamma}
\def\h{\mathfrak{h}}
\def\k{\kappa}
\def\l{\lambda}
\def\o{\omega}
\def\p{\wp}
\def\r{\rho}
\def\t{\tau}
\def\s{\sigma}
\def\z{\zeta}
\def\x{\xi}
\def\V={{{\bf\rm{V}}}}
 \def\A{{\cal{A}}}
 \def\B{{\cal{B}}}
 \def\C{{\cal{C}}}
 \def\D{{\cal{D}}}
\def\G{\Gamma}
\def\K{{\cal{K}}}
\def\O{\Omega}
\def\R{\bar{R}}
\def\T{{\cal{T}}}
\def\L{\Lambda}
\def\f{E_{\tau,\eta}(sl_2)}
\def\E{E_{\tau,\eta}(sl_n)}
\def\Zb{\mathbb{Z}}
\def\Cb{\mathbb{C}}

\def\R{\overline{R}}
% Shorthands for \begin{equation} and the like

\def\beq{\begin{equation}}
\def\eeq{\end{equation}}
\def\bea{\begin{eqnarray}}
\def\eea{\end{eqnarray}}
\def\ba{\begin{array}}
\def\ea{\end{array}}
\def\no{\nonumber}
\def\le{\langle}
\def\re{\rangle}
\def\lt{\left}
\def\rt{\right}

\newtheorem{Theorem}{Theorem}
\newtheorem{Definition}{Definition}
\newtheorem{Proposition}{Proposition}
\newtheorem{Lemma}{Lemma}
\newtheorem{Corollary}{Corollary}
\newcommand{\proof}[1]{{\bf Proof. }
        #1\begin{flushright}$\Box$\end{flushright}}

\baselineskip=20pt

%%%%%%%%%%%%%%%%%%%%%%%%%%%%%%%%%%%%%%%%%%%%%%%%%%%%%%%%%%%%
%                                                          %
%  Title page                                              %
%                                                          %
%%%%%%%%%%%%%%%%%%%%%%%%%%%%%%%%%%%%%%%%%%%%%%%%%%%%%%%%%%%%
\newfont{\elevenmib}{cmmib10 scaled\magstep1}
\newcommand{\preprint}{
   \begin{flushleft}
     %\elevenmib Yukawa\, Institute\, Kyoto\\
   \end{flushleft}\vspace{-1.3cm}
   \begin{flushright}\normalsize
  % \sf  YITP-03-53\\
   %  {\tt hep-th/yymmnnn} \\ November 2005
   \end{flushright}}
\newcommand{\Title}[1]{{\baselineskip=26pt
   \begin{center} \Large \bf #1 \\ \ \\ \end{center}}}
\newcommand{\Author}{\begin{center}
   \large \bf
Xin Zhang${}^{a}$,~Yuan-Yuan Li${}^{a}$,~Junpeng Cao${}^{a,b}$,~Wen-Li Yang${}^{c,d}\footnote{Corresponding author:
wlyang@nwu.edu.cn}$,\\~Kangjie Shi${}^c$ and~Yupeng
Wang${}^{a,b}\footnote{Corresponding author: yupeng@iphy.ac.cn}$
 \end{center}}
\newcommand{\Address}{\begin{center}

     ${}^a$Beijing National Laboratory for Condensed Matter
          Physics, Institute of Physics, Chinese Academy of Sciences, Beijing
           100190, China\\
     ${}^b$Collaborative Innovation Center of Quantum Matter, Beijing,
     China\\
     ${}^c$Institute of Modern Physics, Northwest University,
     Xian 710069, China \\
     ${}^d$Beijing Center for Mathematics and Information Interdisciplinary Sciences, Beijing, 100048,  China
   \end{center}}
\newcommand{\Accepted}[1]{\begin{center}
   {\large \sf #1}\\ \vspace{1mm}{\small \sf Accepted for Publication}
   \end{center}}

\preprint
\thispagestyle{empty}
\bigskip\bigskip\bigskip

\Title{Bethe states of the XXZ spin-$\frac12$ chain with arbitrary boundary fields} \Author

\Address
\vspace{1cm}

\begin{abstract}
Based on the inhomogeneous $T-Q$ relation constructed via the
off-diagonal Bethe Ansatz, the Bethe-type eigenstates of the XXZ
spin-$\frac{1}{2}$ chain with arbitrary boundary fields are constructed. It is found that by employing two sets of gauge transformations, proper generators and reference state for constructing Bethe vectors can be obtained respectively. Given an inhomogeneous $T-Q$ relation for an eigenvalue, it is proven that the resulting Bethe state is an eigenstate of the transfer matrix,
 provided that the parameters of the generators satisfy the associated
Bethe Ansatz equations.

\vspace{1truecm} \noindent {\it PACS:} 75.10.Pq, 03.65.Vf, 71.10.Pm

\end{abstract}
\newpage
%%%%%%%%%%%%%%%%%%%%%%%%%%%%%%%%%%%%%%%%%%%%%%%%%%%%%%%%%%%%%%%
%                                                             %
%  1. Introduction                                            %
%                                                             %
%%%%%%%%%%%%%%%%%%%%%%%%%%%%%%%%%%%%%%%%%%%%%%%%%%%%%%%%%%%%%%%
\section{Introduction}
\label{intro} \setcounter{equation}{0}
In this paper we focus on constructing the Bethe-type eigenstates (Bethe states) of the quantum  XXZ spin-$\frac{1}{2}$ chain
with arbitrary boundary fields, defined by the Hamiltonian
\begin{eqnarray}
H &=&\sum_{j=1}^{N-1}\lt\{\sigma_j^x\sigma_{j+1}^x+\sigma_j^y\sigma^y_{j+1}+\cosh\eta\sigma_j^z\sigma_{j+1}^z\rt\}
    +{\vec h}_1\cdot{\vec \sigma}_1+{\vec h}_N\cdot{\vec \sigma}_N\no\\[4pt]
  &=&\sum_{j=1}^{N-1}\lt\{\sigma_j^x\sigma_{j+1}^x+\sigma_j^y\sigma^y_{j+1}+\cosh\eta\sigma_j^z\sigma_{j+1}^z\rt\}+\no\\[4pt]
  &&+\frac{\sinh\eta}{\sinh\alpha_-\cosh\beta_-}(\cosh\alpha_-\sinh\beta_-\sigma_1^z+\cosh\theta_-\sigma_1^x+i\sinh\theta_-\sigma_1^y)\no\\[4pt]
  &&-\frac{\sinh\eta}{\sinh\alpha_+\cosh\beta_+}(\cosh\alpha_+\sinh\beta_+\sigma_N^z-\cosh\theta_+\sigma_N^x-i\sinh\theta_+\sigma_N^y),\label{Hamiltonian}
\end{eqnarray}
where  $\sigma_{j}^{\alpha}$ $(\alpha=x, y, z)$ is the Pauli matrix on site $j$ and along
$\alpha$ direction, and $\alpha_{\pm}$, $\beta_{\pm}$, $\theta_{\pm}$ are the boundary parameters associated with the boundary fields. The model
has played a fundamental role in the study of quantum integrable system \cite{Gau71,Alc87,Skl88} with boundaries. Moreover, it has many applications in
the non-perturbative analysis of quantum systems appearing in string and super-symmetric Yang-Mills (SYM) theories \cite{Bei12}
(and references therein), low-dimensional condensed matter physics \cite{Gia03} and statistical physics \cite{Jde05, Sir09}. However,
the Bethe Ansatz solution of the model for generic values of boundary fields has challenged for many years since Sklyanin's elegant work \cite{Skl88},
and many efforts had been made
\cite{Nep02,Cao03,Yan04-1,Gie05,Jde05,Doi06,Baj06,Yan06,Gal08,Nic12,Bel133} to approach this nontrivial problem.

The off-diagonal Bethe Ansatz (ODBA) provides an efficient method \cite{Cao1} for solving the eigenvalue problem of integrable models with generic integrable boundary conditions.
Several long-standing models \cite{Cao1,Cao2,Cao2-1,Cao3,Li14,Cao14,Hao14} including the XXZ spin-$\frac12$ chain  have since been solved via this method. The central point is to construct a proper $T-Q$ relation \cite{Bax82, Kor93}, which
immediately leads to the Bethe Ansatz solution for the eigenvalues, with an
extra off-diagonal (or inhomogeneous) term  based on their functional
relations. An interesting issue in this
framework is how to retrieve the Bethe states from the obtained spectrum. Indeed, significant progress has been achieved in this aspect recently. For example, based on the inhomogeneous $T-Q$ relation obtained in \cite{Cao2}, the Bethe
states of the open $XXX$ spin chain was conjectured in \cite{Bel13} via the  modified algebraic Bethe ansatz and then proven in \cite{Cao-14-Bethe-state}. Alternatively, a set of eigenstates of the inhomogeneous XXZ transfer matrix was derived in \cite{Nic12,Fad14} via the separation of variables (SoV) method \cite{Skl92}. However, how to get the homogeneous limit (if there is any) of those SoV states  is still an open problem. It is also interesting that the eigenstates in homogeneous limit can be classified by the representation of
the q-Onsager algebra \cite{Bac06,Bac13}.

For the open XXZ chain, when the boundary fields are all along $z$-direction (or the diagonal boundaries), the corresponding
Bethe states were constructed by the algebraic Bethe Ansatz method \cite{Skl88,Skl78}. The unparallel boundary fields break the
$U(1)$-symmetry (i.e, the total spin is not conserved any more). This makes the problem of constructing Bethe vectors rather unusual because of the absence of an obvious reference state.
So far, the Bethe states could only be obtained for some constrained boundary parameters. When the boundary parameters obey a
constraint \cite{Nep02,Cao03}, which is already in $U(1)$-symmetry-broken case, the associated
Bethe states were constructed \cite{Cao03} within the framework of the generalized algebraic Bethe Ansatz
\cite{Bax82,Tak79}. Very recently, based on  the inhomogeneous $T-Q$ relation  and small sites analysis of the model with triangular boundaries,
the corresponding Bethe states  are conjectured \cite{Bel14}  and proven in \cite{Cra14}. In this paper we study the Bethe states of the transfer matrix for
the quantum  XXZ spin-$\frac{1}{2}$ chain with arbitrary boundary fields based on the inhomogeneous $T-Q$ relation of the
eigenvalues obtained by ODBA.

The paper is organized as follows. Section 2 serves as an
introduction of our notations and the ODBA solutions of the model. In section 3, after introducing the gauge transformations and the
associated left (right)  state,
we compute the associated commutation relations
among the matrix elements of the two gauged double-row monodromy matrices, and their actions on the
associated  state.  In section 4, two particular gauge transformations are chosen according to
the boundary parameters of $K$-matrices respectively. Based on the chosen  parameters of the resulting transformations,
the Bethe-type eigenstates of the transfer matrix are constructed.  In section 5, we summarize our
results and give the concluding remarks. Some useful formulae and technical proofs are
given in Appendices A-C respectively.

%%%%%%%%%%%%%%%%%%%%%%%%%%%%%%%%%%%%%%%%%%%%%%%%%%%%%%%%%%%%%%%
%                                                             %
%  2. Transfer matrix  and eigenvalues                        %
%                                                             %
%                                                             %
%                                                             %
%%%%%%%%%%%%%%%%%%%%%%%%%%%%%%%%%%%%%%%%%%%%%%%%%%%%%%%%%%%%%%%

\section{ODBA solution}
\label{ODBA} \setcounter{equation}{0}

Let ${V}$ be a two-dimensional vector space. For the XXZ spin
chain with generic boundaries, the associated $R$-matrix  and the  reflection matrices $K^{\mp}(u)$ \cite{Veg93,Gho94} read
\begin{eqnarray}
R(u)=\frac{1}{\sinh\eta}\left(
                          \begin{array}{cccc}
                            \sinh(u+\eta) & 0 & 0 & 0 \\
                            0 & \sinh u & \sinh\eta & 0 \\
                            0 & \sinh\eta & \sinh u & 0 \\
                            0 & 0 & 0 & \sinh(u+\eta) \\
                          \end{array}
                        \right),\label{def R matrix}
\end{eqnarray}
 \bea
K^-(u)&=&\lt(\begin{array}{ll}K^-_{11}(u)&K^-_{12}(u)\\
K^-_{21}(u)&K^-_{22}(u)\end{array}\rt),\no\\
K^-_{11}(u)&=&2\lt(\sinh(\a_-)\cosh(\b_{-})\cosh(u)
+\cosh(\a_-)\sinh(\b_-)\sinh(u)\rt),\no\\
K^-_{22}(u)&=&2\lt(\sinh(\a_-)\cosh(\b_{-})\cosh(u)
-\cosh(\a_-)\sinh(\b_-)\sinh(u)\rt),\no\\
K^-_{12}(u)&=&e^{\theta_-}\sinh(2u),\quad
K^-_{21}(u)=e^{-\theta_-}\sinh(2u),\label{K-matrix}\eea and \bea
K^+(u)=\lt.K^-(-u-\eta)\rt|_{(\a_-,\b_-,\theta_-)\rightarrow
(-\a_+,-\b_+,\theta_+)},\label{K-6-2} \eea
where $\eta$ is the crossing parameter, and $\a_{\mp},\,\b_{\mp},\,\theta_{\mp}$ are the boundary parameters
associated with boundary fields (see (\ref{Hamiltonian})). The $R$-matrix is a
solution of the quantum Yang-Baxter equation (QYBE) \bea
R_{12}(u_1-u_2)R_{13}(u_1-u_3)R_{23}(u_2-u_3)
=R_{23}(u_2-u_3)R_{13}(u_1-u_3)R_{12}(u_1-u_2),\label{QYB}
\eea
and $K^{\mp}(u)$ satisfy the following reflection equations (RE)
 \bea &&R_{12}(u_1-u_2)K^-_1(u_1)R_{21}(u_1+u_2)K^-_2(u_2)\no\\
 &&~~~~~~=
K^-_2(u_2)R_{12}(u_1+u_2)K^-_1(u_1)R_{21}(u_1-u_2),\label{RE-V}\eea
and \bea
&&R_{12}(u_2-u_1)K^+_1(u_1)R_{21}(-u_1-u_2-2\eta)K^+_2(u_2)\no\\
&&~~~~~~= K^+_2(u_2)R_{12}(-u_1-u_2-2\eta)K^+_1(u_1)R_{21}(u_2-u_1).
\label{DRE-V}\eea Here
and below we adopt the standard notations: for any matrix $A\in {\rm
End}({ V})$, $A_j$ is an embedding operator in the tensor
space ${ V}\otimes { V}\otimes\cdots$, which acts as $A$ on the
$j$-th space and as identity on the other factor spaces; $R_{ij}(u)$
is an embedding operator of $R$-matrix in the tensor space, which
acts as identity on the factor spaces except for the $i$-th and
$j$-th ones.

We introduce the ``row-to-row"  (or one-row ) monodromy matrices
$T_0(u)$ and $\hat{T}_0(u)$, which are $2\times 2$ matrices with
elements being operators acting on the tensor space ${V}^{\otimes N}$, \bea
T_0(u)&=&R_{0N}(u-\theta_N)R_{0\,N-1}(u-\theta_{N-1})\cdots
R_{01}(u-\theta_1),\label{Mon-V-1}\\
\hat{T}_0(u)&=&R_{10}(u+\theta_1)R_{20}(u+\theta_{2})\cdots
R_{N0}(u+\theta_N).\label{Mon-V-2} \eea Here
$\{\theta_j|j=1,\cdots,N\}$ are the inhomogeneous parameters. For open spin chains,
one needs to consider  the double-row monodromy matrix $\mathscr{U}_0(u)$
\bea
  \mathscr{U}_0(u)=T_0(u)K^-_0(u)\hat{T}_0(u).
  \label{Mon-V-0}
\eea The double-row transfer matrix $t(u)$  is thus given by \bea
t(u)=tr_0(K^+_0(u)\mathscr{U}_0(u)).\label{trans}\eea The QYBE
(\ref{QYB}) and  REs (\ref{RE-V}) and (\ref{DRE-V}) lead to
the fact that the transfer matrices with different spectral
parameters commute with each other \cite{Skl88}: $[t(u),t(v)]=0$.
Then $t(u)$ serves as the generating functional of the conserved
quantities of the corresponding system, which ensures the
integrability of the open spin chain.

The Hamiltonian (\ref{Hamiltonian}) is expressed in
terms of the transfer matrix (\ref{trans}) with the $K$-matrices (\ref{K-matrix}) and (\ref{K-6-2}) by
\begin{eqnarray}
&&H=\sinh\eta \frac{\partial \ln t(u)}{\partial
u}|_{u=0,\theta_j=0}-N\cosh\eta -\tanh\eta\sinh\eta.
\eea

It was proven in \cite{Cao2-1} that for generic $\{\theta_j\}$ the transfer matrix given by (\ref{trans})
for arbitrary boundary parameters
satisfies the following operator identities
\bea
t(\theta_j)\,t(\theta_j-\eta)&=&a(\theta_j)d(\theta_j-\eta)\times {\rm id},
\label{Operator-id-3}\\
t(-u-\eta)&=&t(u),\quad t(u+i\pi)=t(u),\label{Periodic-6}\\
t(0)&=&-2^3\sinh\a_-\cosh\b_-\sinh\a_+\cosh\b_+\cosh\eta\,\no\\
&&\times
\prod_{l=1}^N\frac{\sinh(\eta-\theta_l)\,\sinh(\eta+\theta_l)}{\sinh^2\eta}\times
{\rm id},\\
t(\frac{i\pi}{2})&=&-2^3\cosh\a_-\sinh\b_-\cosh\a_+\sinh\b_+\cosh\eta\,\no\\
&&\times
\prod_{l=1}^N\frac{\sinh(\frac{i\pi}{2}+\theta_l+\eta)\sinh(\frac{i\pi}{2}+\theta_l-\eta)}{\sinh^2\eta}
\times {\rm id},\\
\lim_{u\rightarrow \pm\infty}
t(u)&=&-\frac{\cosh(\theta_--\theta_+)e^{\pm[(2N+4)u+(N+2)\eta]}}
{2^{2N+1}\sinh^{2N}\eta}\times {\rm id} +\ldots,\label{Tran-Asy-6}
\eea
where the functions $a(u)$ and $d(u)$ are  given by
\bea
a(u)&=&\hspace{-0.32truecm}-\hspace{-0.06truecm}2^2
\frac{\sinh(2u\hspace{-0.08truecm}+\hspace{-0.08truecm}2\eta)}{\sinh(2u\hspace{-0.08truecm}+\hspace{-0.08truecm}\eta)}
\sinh(u\hspace{-0.08truecm}-\hspace{-0.08truecm}\a_-)\cosh(u\hspace{-0.08truecm}-\hspace{-0.08truecm}\b_-)
\sinh(u\hspace{-0.08truecm}-\hspace{-0.08truecm}\a_+)
\cosh(u\hspace{-0.08truecm}-\hspace{-0.08truecm}\b_+)\bar{A}(u),\label{bar-a}\\
d(u)&=&a(-u-\eta),\quad \bar{A}(u)=
\prod_{l=1}^N\frac{\sinh(u-\theta_l+\eta)\sinh(u+\theta_l+\eta)}
{\sinh^2\eta}. \label{A-function} \eea
The above operator relations lead to that the corresponding eigenvalue of the transfer
matrix, denoted by $\Lambda(u)$, enjoys the following properties
\bea
&&\hspace{-1.2truecm}\Lambda(\theta_j)\Lambda(\theta_j-\eta)=a(\theta_j)d(\theta_j-\eta),
\quad j=1,\ldots,N, \label{Eigen-Identity-6-1}\\[2pt]
&&\hspace{-1.2truecm}\L(-u-\eta)=\L(u),\quad \L(u+i\pi)=\L(u),\label{crosing-Eign-6}\\[2pt]
&&\hspace{-1.2truecm}\L(0)=-2^3\sinh\a_-\cosh\b_-\sinh\a_+\cosh\b_+\cosh\eta\,
\prod_{l=1}^N\frac{\sinh(\eta-\theta_l)\,\sinh(\eta+\theta_l)}{\sinh^2\eta},\label{Eigen-6-1}\\[2pt]
&&\hspace{-1.2truecm}\L(\frac{i\pi}{2})=-2^3\cosh\a_-\sinh\b_-\cosh\a_+\sinh\b_+\cosh\eta\,\no\\[2pt]
&&\hspace{-1.2truecm}\quad\quad\quad\quad \times \prod_{l=1}^N\frac{\sinh(\frac{i\pi}{2}+\theta_l+\eta)\,\sinh(\frac{i\pi}{2}+\theta_l-\eta)}{\sinh^2\eta},\label{Eigen-6-2}\\[2pt]
&&\hspace{-1.2truecm}\lim_{u\rightarrow \pm\infty}\L(u)=-\frac{\cosh(\theta_--\theta_+)e^{\pm[(2N+4)u+(N+2)\eta]}}
{2^{2N+1}\sinh^{2N}\eta}+\ldots.\label{Eigen-Asy-6}\\[2pt]
&&\hspace{-1.2truecm}\L(u) \mbox{, as an entire function of $u$, is a
trigonometric polynomial of degree $2N+4$}.\label{Eigen-Anal-6}
\eea
Each solution of (\ref{Eigen-Identity-6-1})-(\ref{Eigen-Anal-6}) can be  given in terms of the following
inhomogeneous $T-Q$ relation \cite{Cao2,Cao2-1,Nep13,Cao14-2} \footnote{The inhomogeneous $T-Q$ relation (\ref{T-Q-1}) corresponds to the  special case (i.e., $M=0$) of the general ones in \cite{Cao2}, which was first proposed  for the $XXX$ case and its validity for the XXZ case was also pointed out in \cite{Cao2-1}.
The relation was then confirmed by the SoV method \cite{Kit14} for the $XXZ$ case and its generalization to higher spin case was given in \cite{Cao14-2,Cao14-3}.  }
\bea
 \Lambda(u)&=&a(u)\frac{Q(u-\eta)}{Q(u)}+d(u)\frac{Q(u+\eta)}{Q(u)}\no\\
 &&+\frac{2c\,\sinh(2u)\sinh(2u+2\eta)}{Q(u)}\bar A(u)\bar A(-u-\eta),\label{T-Q-1}
\eea where $c$ is a constant depending on the boundary parameters
\bea
c=\cosh(\a_-+\b_-+\a_++\b_++(1+N)\eta)-\cosh(\theta_--\theta_+),\label{c-constant}
\eea
and the $Q$-function is given by
\bea
Q(u)=\prod_{j=1}^N\frac{\sinh(u-\l_j)\sinh(u+\l_j+\eta)}{\sinh\eta\,\sinh\eta},\label{Q-function-1}
\eea with the parameters $\{\l_j\}$ satisfying the associated Bethe ansatz equations (BAEs)
\bea
a(\l_j)Q(\l_j-\eta)+d(\l_j)Q(\l_j+\eta)&+&2c\,\sinh2\l_j\sinh(2\l_j+2\eta)\bar{A}(\l_j)\bar{A}(-\l_j-\eta)=0,\no\\
j&=&1,\ldots,N.\label{BAE-1}
\eea
We shall show in Section 4 that for each solution of (\ref{Eigen-Identity-6-1})-(\ref{Eigen-Anal-6}), one can construct a corresponding Bethe-type eigenstate
(see  (\ref{Bethe-right}) below) of the transfer
matrix (\ref{trans}) with the eigenvalue given by (\ref{T-Q-1}). Therefore the relations (\ref{Eigen-Identity-6-1})-(\ref{Eigen-Anal-6}) (or the inhomogeneous $T-Q$ relation (\ref{T-Q-1})) indeed completely characterize the spectrum of the transfer matrix.

Some remarks are in order.  There exist
various  possible ways \cite{Cao2} to parameterize  the
solution of (\ref{Eigen-Identity-6-1})-(\ref{Eigen-Anal-6}), but they are all equivalent to each other
because of the finite number of solutions. For generic boundary parameters, the minimal degree of the
$Q$-polynomial is $N$, while the degree of the
$Q$-polynomial may be reduced to a small value in case of the inhomogeneous term (or the third term in (\ref{T-Q-1})) vanishing.
In this case the $T-Q$ relation becomes a homogeneous one (the well-known Baxter's $T-Q$ relation). This happens in
case of $U(1)$ symmetry or in degenerate cases \cite{Cao03}, for which the transfer matrix can be
diagonalized in smaller blocks.

%%%%%%%%%%%%%%%%%%%%%%%%%%%%%%%%%%%%%%%%%%%%%%%%%%%%%%%%%%%%%%%
%                                                             %
%  4. Gauge transformation                                     %
%                                                             %
%                                                             %
%                                                             %
%%%%%%%%%%%%%%%%%%%%%%%%%%%%%%%%%%%%%%%%%%%%%%%%%%%%%%%%%%%%%%%

\section{Gauge transformations and the associated operators}
\label{Gauge} \setcounter{equation}{0}

A particular set of gauge transformation (the six-vertex version of the vertex-face correspondence),
which have played a key role to construct the associated Bethe states,  was proposed in \cite{Cao03}. Recently, such gauge transformation was adopted in constructing the  SoV eigenstates \cite{Fad14} and the Bethe states \cite{Bel14} for the open chains.
In this paper, we use two sets of such gauge transformation and
the inhomogeneous $T-Q$ relation (\ref{T-Q-1}) to construct the Bethe states for the quantum  XXZ spin-$\frac{1}{2}$ chain with arbitrary boundary fields.

Following \cite{Cao03}, let us introduce two column vectors as follows
\begin{eqnarray}
X_m(u|\a)=\left(
         \begin{array}{c}
           e^{-[u+(\alpha+m)\eta]} \\
           1 \\
         \end{array}
       \right),\quad\quad
Y_m(u|\a)=\left(
         \begin{array}{c}
           e^{-[u+(\alpha-m)\eta]} \\
           1 \\
         \end{array}
       \right),\label{Intertwiner-1}
\end{eqnarray}
where $\alpha$ and $m$ are two arbitrary complex parameters. For generic $\a$ and $m$, the two vectors
are linearly independent. Thus one can introduce the following gauge matrices
\begin{eqnarray}
&&\overline{M}_m(u|\a)=\left(
                    \begin{array}{cc}
                      X_{m}(u|\a), & Y_m(u|\a) \\
                    \end{array}
                  \right),\quad
\overline{M}^{-1}_m(u)=\left(
                          \begin{array}{c}
                            \overline{Y}_m(u|\a) \\
                            \overline{X}_m(u|\a) \\
                          \end{array}
                        \right),\label{def M}\\[4pt]
&&\widetilde{M}_m(u|\a)=\left(
                    \begin{array}{cc}
                      X_{m+1}(u|\a), & Y_{m-1}(u|\a) \\
                    \end{array}
                  \right),\quad
\widetilde{M}^{-1}_m(u|\a)=\left(
                          \begin{array}{c}
                            \widetilde{Y}_{m-1}(u|\a) \\
                            \widetilde{X}_{m+1}(u|\a) \\
                          \end{array}
                        \right),\label{def tilde M}\\[4pt]
&&\widehat{M}_m(u|\a)=\left(
                     \begin{array}{cc}
                       \widehat{X}_{m-1}(u|\a), & \widehat{Y}_{m+1}(u|\a) \\
                     \end{array}
                   \right),\quad
\widehat{M}_m^{-1}(u|\a)=\left(
                        \begin{array}{c}
                          \overline{Y}_{m+1}(u|\a) \\
                          \overline{X}_{m-1}(u|\a) \\
                        \end{array}
                      \right),\label{def hat M}
\end{eqnarray}
where
\begin{eqnarray}
&&\overline{X}_m(u|\a)=\frac{e^{u+\alpha\eta}}{2\sinh m\eta}\left(
                                                      \begin{array}{cc}
                                                        1 ,& -e^{-[u+(\alpha+m)\eta]} \\
                                                      \end{array}
                                                    \right),\label{Over-X}\\[4pt]
&&\overline{Y}_m(u|\a)=\frac{e^{u+\alpha\eta}}{2\sinh m\eta}\left(
                                                      \begin{array}{cc}
                                                        -1 ,& e^{-[u+(\alpha-m)\eta]} \\
                                                      \end{array}
                                                    \right),\\[4pt]
&&\widetilde{X}_m(u|\a)=\frac{e^{\eta}\sinh m\eta}{\sinh(m-1)\eta}\overline{X}_{m}(u|\a),\quad
\widetilde{Y}_m(u|\a)=\frac{e^{\eta}\sinh m\eta}{\sinh (m+1)\eta}\overline{Y}_{m}(u|\a),\\[4pt]
&&\widehat{X}_m(u|\a)=\frac{e^{-\eta}\sinh(m+2)\eta}{\sinh(m+1)\eta}X_m(u|\a),\,\,
\widehat{Y}_m(u|\a)=\frac{e^{-\eta}\sinh(m\hspace{-0.04truecm}-\hspace{-0.04truecm}2)\eta}
{\sinh(m\hspace{-0.04truecm}-\hspace{-0.04truecm}1)\eta}Y_{m}(u|\a).\label{Intertwiner-2}
\end{eqnarray}  We remark that the vectors $X_m(u|\a)$ and $\overline{X}_m(u|\a)$ only depend on $\a+m$, while the vectors
 $Y_m(u|\a)$ and $\overline{Y}_m(u|\a)$ only depend on $\a-m$, up to a scaling factor.

These column and row vectors satisfy some intertwining relations \cite{Cao03}, which
are listed in Appendix A (see (\ref{relation1})-(\ref{relation28}) below). These relations allow us
to introduce the following gauged operators and the associated
$K^+$-matrix
\begin{eqnarray}
\hspace{-0.8truecm} \overline{\mathscr{U}}(m,\a|u)&=&\left(
                      \begin{array}{cc}
                        \overline{\mathscr{A}}_m(u|\a) & \overline{\mathscr{B}}_m(u|\a) \\[2pt]
                        \overline{\mathscr{C}}_m(u|\a) & \overline{\mathscr{D}}_m(u|\a) \\
                      \end{array}
                    \right)\no\\[4pt]
                    &=&
 \left(
    \begin{array}{cc}
      \overline{Y}_m(u|\a)\mathscr{U}(u)\widehat{X}_{m-2}(-u|\a) & \overline{Y}_{m}(u|\a)\mathscr{U}(u)\widehat{Y}_{m}(-u|\a) \\[2pt]
      \overline{X}_m(u|\a)\mathscr{U}(u)\widehat{X}_{m}(-u|\a) & \overline{X}_{m}(u|\a)\mathscr{U}(u)\widehat{Y}_{m+2}(-u|\a) \\
    \end{array}
  \right),\label{def U left}\\[4pt]
\hspace{-0.8truecm}\overline{K}^+(m,\a|u)&=&\left(
                     \begin{array}{cc}
                       {\overline{K}^+_{11}}(m,\a|u) & {\overline{K}^+_{12}}(m,\a|u) \\[2pt]
                       {\overline{K}^+_{21}}(m,\a|u) & {\overline{K}^+_{22}}(m,\a|u) \\
                     \end{array}
                   \right)\no\\[4pt]
&=&\left(
   \begin{array}{cc}
     \overline{Y}_{m}(-u|\a)K^+(u)X_m(u|\a) & \overline{Y}_{m+2}(-u|\a)K^+(u)Y_m(u|\a) \\
     \overline{X}_{m-2}(-u|\a)K^+(u)X_{m}(u|\a) & \overline{X}_{m}(-u|\a)K^+(u)Y_{m}(u|\a) \\
   \end{array}
 \right).\label{def K+ left}
\end{eqnarray}
With the help of the relations (\ref{orth M})-(\ref{orth hat M}), we can rewrite the transfer matrix (\ref{trans}) in terms of
the above gauged operators and $K$-matrix, namely,
\bea
t(u)&=&{\rm{tr}}\lt\{{K^+}(u)\mathscr{U}(u)\rt\}\no\\
&=&\overline{K}^+_{11}(m,\a|u)\overline{\mathscr{A}}_m(u|\a)+\overline{K}^+_{21}(m,\a|u)\overline{\mathscr{B}}_m(u|\a)\no\\
&&+\overline{K}^+_{12}(m,\a|u)\overline{\mathscr{C}}_m(u|\a)+\overline{K}^+_{22}(m,\a|u)\overline{\mathscr{D}}_m(u|\a)\no\\
&=&{\rm{tr}}\lt\{\overline{\mathscr{U}}(m,\a|u)\overline{K}^+(m,\a|u)\rt\}.\label{trans-1}
\eea

The QYBE (\ref{QYB}), the RE (\ref{RE-V}) and the intertwining relations given in Appendix A allow us to derive the commutation relations
among the matrix elements of $\overline{\mathscr{U}}(m,\a|u)$. Here we present some relevant relations for our purpose:
 \begin{eqnarray}
&& \overline{\mathscr{C}}_m(u_1|\a)\overline{\mathscr{C}}_{m+2}(u_2|\a)=\overline{\mathscr{C}}_m(u_2|\a)
 \overline{\mathscr{C}}_{m+2}(u_1|\a),\label{CC relation left}\\[4pt]
&&\lt[\overline{\mathscr{D}}_{m-2}(u_2|\a),\,\overline{\mathscr{D}}_{m-2}(u_1|\a)\rt]=
\frac{\sinh(m\eta+u_1+u_2)\sinh\eta}{\sinh m\eta\sinh(u_1+u_2+\eta)}\overline{\mathscr{C}}_{m-2}(u_1|\a)\overline{\mathscr{B}}_{m}(u_2|\a)\no\\[2pt]
&&\qquad-\frac{\sinh(m\eta+u_1+u_2)\sinh\eta}{\sinh m\eta\sinh(u_1+u_2+\eta)}\overline{\mathscr{C}}_{m-2}(u_2|\a)\overline{\mathscr{B}}_{m}(u_1|\a),
\label{DD relation left}\\[4pt]
&&\overline{\mathscr{D}}_{m-2}(u_2|\a)\overline{\mathscr{C}}_{m-2}(u_1|\a)=\frac{\sinh(u_1-u_2+\eta)\sinh(u_1+u_2)}{\sinh(u_1+u_2+\eta)\sinh(u_1-u_2)}
\overline{\mathscr{C}}_{m-2}(u_1|\a)\overline{\mathscr{D}}_{m}(u_2|\a)\no\\[2pt]
&&\qquad-\frac{\sinh(m\eta-u_1+u_2)\sinh(u_1+u_2)\sinh\eta}{\sinh m\eta\sinh(u_1-u_2)\sinh(u_1+u_2+\eta)}
\overline{\mathscr{C}}_{m-2}(u_2|\a)\overline{\mathscr{D}}_{m}(u_1|\a)\no\\[2pt]
&&\qquad-\frac{\sinh(m\eta+u_1+u_2)\sinh\eta}{\sinh m\eta\sinh(u_1+u_2+\eta)}
\overline{\mathscr{C}}_{m-2}(u_2|\a)\overline{\mathscr{A}}_{m}(u_1|\a),\label{DC relation left}\\[4pt]
&&\lt[\overline{\mathscr{D}}_m(u_2|\a),\, \overline{\mathscr{A}}_m(u_1|\a)\rt]=
\frac{\sinh(m+1)\eta\sinh\eta\sinh(m\eta-u_1+u_2)\sinh(u_1+u_2+2\eta)}{\sinh(m+2)\eta\sinh(m-1)\eta\sinh(u_1-u_2)\sinh(u_1+u_2+\eta)}\no\\[2pt]
&&\qquad\times[\overline{\mathscr{C}}_m(u_1|\a)\overline{\mathscr{B}}_{m+2}(u_2|\a)-
\overline{\mathscr{C}}_m(u_2|\a)\overline{\mathscr{B}}_{m+2}(u_1|\a)].\label{DA relation 2}
\end{eqnarray} The proof of the above relations is relegated to Appendix B.

Let us introduce the following left local  states of the $n$-th site in
the lattice:
\begin{eqnarray}
\langle{\omega};m,\a|_n=\overline{X}_{m+n-N-1}(\theta_n|\a),\quad n=1,\cdots,N, \label{Local-left-Vacuum}
\end{eqnarray}  where the row vector $\overline{X}_m(u)$ is given by (\ref{Over-X}). Further, we introduce
the following global  state from the above local states,
\begin{eqnarray}
\langle{\a+m}|&=&2^Ne^{-\sum_{l=1}^N\theta_l-\a N\eta}\prod_{l=1}^N\sinh(m-l)\eta\,\bigotimes_{n=1}^N\langle{\omega};m,\a|_n.\label{Left-vacuum}
\end{eqnarray}
The explicit expression (\ref{Over-X}) of the row
vector $\overline{X}_m(u)$ implies that the above state  does only depend  on $\a+m$.  Following the method in \cite{Cao03,Fan96,Yan04}, after some tedious calculation, we obtain the
actions of the gauged operators $\overline{\mathscr{C}}_m(u|\a)$, $\overline{\mathscr{A}}_m(u|\a)$ and $\overline{\mathscr{D}}_m(u|\a)$ on the state (\ref{Left-vacuum}) as follows:
\begin{eqnarray}
\hspace{-1.2truecm}\langle{\a+m}|\overline{\mathscr{C}}_m(u|\a)&=&
   \hspace{-.32truecm}{\overline{K}^-_{21}}(m-N,\a|u)\frac{\sinh(m+2)\eta}{\sinh(m+2-N)\eta}\no\\
   &&\times\prod_{j=1}^N\frac{\sinh(u-\theta_j+\eta)\sinh(u+\theta_j)}{\sinh^2\eta}\langle{\a+m+2}|,\label{C left function}\\
\hspace{-1.2truecm}\langle{\a+m}|\overline{\mathscr{D}}_m(u|\a)&=&
   \hspace{-0.32truecm}{\overline{K}^-_{22}}(m\hspace{-0.04truecm}-\hspace{-0.04truecm}N,\a|u)\prod_{j=1}^N
   \frac{\sinh(u\hspace{-0.04truecm}-\hspace{-0.04truecm}\theta_j\hspace{-0.04truecm}+\hspace{-0.04truecm}\eta)
   \sinh(u\hspace{-0.04truecm}+\hspace{-0.04truecm}\theta_j\hspace{-0.04truecm}+\hspace{-0.04truecm}\eta)}
   {\sinh^2\eta}\langle{\a\hspace{-0.04truecm}+\hspace{-0.04truecm}m}|\no\\
   &&\hspace{-0.32truecm}+{\overline{K}^-_{21}}(m\hspace{-0.04truecm}-\hspace{-0.04truecm}N,\a|u)
   \prod_{j=1}^N\frac{\sinh(u\hspace{-0.04truecm}-\hspace{-0.04truecm}\theta_j\hspace{-0.04truecm}+\hspace{-0.04truecm}\eta)}
   {\sinh\eta}\langle{\a\hspace{-0.04truecm}+\hspace{-0.04truecm}m\hspace{-0.04truecm}+\hspace{-0.04truecm}1}|
   \overline{{B}}_{m+1}(u|\a),\label{D left funciton}\\
\hspace{-1.2truecm}\langle{\a+m}|\overline{\mathscr{A}}_m(u|\a)&=&\frac{\sinh(2u-(m-1)\eta)\sinh\eta}{\sinh(2u+\eta)\sinh(1-m)\eta}\no\\
&&\hspace{-0.32truecm}\times\lt\{{\overline{K}^-_{22}}(m\hspace{-0.04truecm}-\hspace{-0.04truecm}N,\a|u)\prod_{j=1}^N
   \frac{\sinh(u\hspace{-0.04truecm}-\hspace{-0.04truecm}\theta_j\hspace{-0.04truecm}+\hspace{-0.04truecm}\eta)
   \sinh(u\hspace{-0.04truecm}+\hspace{-0.04truecm}\theta_j\hspace{-0.04truecm}+\hspace{-0.04truecm}\eta)}
   {\sinh^2\eta}\langle{\a\hspace{-0.04truecm}+\hspace{-0.04truecm}m}|\rt.\no\\
   &&\quad+\lt.{\overline{K}^-_{21}}(m\hspace{-0.04truecm}-\hspace{-0.04truecm}N,\a|u)
   \prod_{j=1}^N\frac{\sinh(u\hspace{-0.04truecm}-\hspace{-0.04truecm}\theta_j\hspace{-0.04truecm}+\hspace{-0.04truecm}\eta)}
   {\sinh\eta}\langle{\a\hspace{-0.04truecm}+\hspace{-0.04truecm}m\hspace{-0.04truecm}+\hspace{-0.04truecm}1}|
   \overline{{B}}_{m+1}(u|\a)\rt\}\no\\
&&+F(u).\label{A left funciton}
\end{eqnarray} Here we have introduced  the gauged $K^-$-matrix
\begin{eqnarray}
{\overline{K}^-}(l',\a|u)&=&\left(
                     \begin{array}{cc}
                       {\overline{K}^-_{11}}(l',\a|u) & {\overline{K}^-_{12}}(l',\a|u) \\
                       {\overline{K}^-_{21}}(l',\a|u) & {\overline{K}^-_{22}}(l',\a|u) \\
                     \end{array}
                   \right)\no\\[4pt]
&=&\left(
   \begin{array}{cc}
     \overline{Y}_{l'}(u|\a)K^-(u)\widehat{X}_{l'-2}(-u|\a) & \overline{Y}_{l'}(u|\a)K^-(u)\widehat{Y}_{l'}(-u|\a) \\
     \overline{X}_{l'}(u|\a)K^-(u)\widehat{X}_{l'}(-u|\a) & \overline{X}_{l'}(u|\a)K^-(u)\widehat{Y}_{l'+2}(-u|\a) \\
   \end{array}
 \right),\label{def K- left}
\end{eqnarray} with  $l'=m-N$, and the gauged operator $\overline{{B}}_{m}(u|\a)$ is given by
\bea
\overline{{B}}_{m}(u|\a)=\overline{Y}_{m-N+1}(-u|\a)\hat{T}(u)\widehat{Y}_{m+1}(-u|\a).
\eea
The extra term $F(u)$ in (\ref{A left funciton}) actually vanishes at the points $\{-\theta_j|j=1,\cdots,N\}$, namely,
\bea
F(-\theta_j)=0,\quad j=1,\ldots,N.\label{Vanishing}
\eea This fact gives rise  to the following important relations
\begin{eqnarray}
\langle{\a+m}|\overline{\mathscr{A}}_m(-\theta_j|\a)=-\frac{\sinh((m-1)\eta+2\theta_j)\sinh\eta}{\sinh(m-1)\eta\sinh(2\theta_j-\eta)}
\langle{\a+m}|\overline{\mathscr{D}}_m(-\theta_j|\a).\label{DA relation special}
\end{eqnarray}

The associated right state (c.f. (\ref{Left-vacuum})), which only depends on $\a+m$, is given by \cite{Cao03}
\begin{eqnarray}
|\a+m\rangle=\bigotimes_{n=1}^NX_{m+N-n+1}(\theta_n|\a),\label{Right-vacumm}
\end{eqnarray} and the associated gauged operators are
\bea
\mathscr{U}(m,\a|u)&=&\left(
                     \begin{array}{cc}
                       \mathscr{A}_m(u|\a) & \mathscr{B}_m(u|\a) \\
                       \mathscr{C}_m(u|\a) & \mathscr{D}_m(u|\a) \\
                     \end{array}
                   \right),\no\\[2pt]
                   &=&\left(
                        \begin{array}{cc}
                          \widetilde{Y}_{m-2}(u|\a)\mathscr{U}(u)X_m(-u|\a) & \widetilde{Y}_{m}(u|\a)\mathscr{U}(u)Y_m(-u|\a) \\[4pt]
                          \widetilde{X}_{m}(u|\a)\mathscr{U}(u)X_m(-u|\a) & \widetilde{X}_{m+2}(u|\a)\mathscr{U}(u)Y_m(-u|\a) \\
                        \end{array}
                      \right).\label{def U+ right}
\eea The matrix elements of the above gauged monodromy matrix acting on the state (\ref{Right-vacumm}) were given in \cite{Cao03}.
Here we present some relevant ones
\begin{eqnarray}
\mathscr{C}_m(u|\a)|\a+m\rangle&=&{K^-_{21}}(l,\a|u)\frac{\sinh(m+N-1)\eta}{\sinh(m-1)\eta}\no\\
&&\times\prod_{j=1}^N\frac{\sinh(u-\theta_j)\sinh(u+\theta_j+\eta)}{\sinh^2\eta}|\a+m-2\rangle,\label{C right function}\\
\mathscr{A}_m(u|\a)|\a+m\rangle&=&{K^-_{11}}(l,\a|u)\prod_{j=1}^N\frac{\sinh(u-\theta_j+\eta)\sinh(u+\theta_j+\eta)}{\sinh^2\eta}|\a+m\rangle\no\\
&&+{K^-_{21}}(l,\a|u)\prod_{j=1}^N\frac{\sinh(u+\theta_j+\eta)}{\sinh\eta}B_{m-1}(u|\a)|\a+m-1\rangle,\label{A right function}
\end{eqnarray}
with $l=m+N$. Here another gauged $K^-$-matrix is (c.f., (\ref{def K- left}))
\begin{eqnarray}
{K^-}(l,\a|u)&=&\left(
                 \begin{array}{cc}
                   {K^-_{11}}(l,\a|u) & {K^-_{12}}(l,\a|u) \\[2pt]
                   {K^-_{21}}(l,\a|u) & {K^-_{22}}(l,\a|u) \\
                 \end{array}
               \right),\no\\[4pt]
               &=&\left(
                    \begin{array}{cc}
                      \widetilde{Y}_{l-2}(u|\a)K^-(u)X_{l}(-u|\a) & \widetilde{Y}_{l}(u|\a)K^-(u)Y_{l}(-u|\a) \\[2pt]
                      \widetilde{X}_{l}(u|\a)K^-(u)X_{l}(-u|\a) & \widetilde{X}_{l+2}(u|\a)K^-(u)Y_{l}(-u|\a) \\
                    \end{array}
                  \right),\label{def K- right}
\end{eqnarray} and the gauged operator $B_{m}(u|\a)$ is given by
\bea
B_{m}(u|\a)=\widetilde{Y}_{m-1}(u)T(u)Y_{m+N-1}(u).
\eea

%%%%%%%%%%%%%%%%%%%%%%%%%%%%%%%%%%%%%%%%%%%%%%%%%%%%%%%%%%%%%%%
%                                                             %
%  5.Bethe states                                             %
%                                                             %
%                                                             %
%                                                             %
%%%%%%%%%%%%%%%%%%%%%%%%%%%%%%%%%%%%%%%%%%%%%%%%%%%%%%%%%%%%%%%

\section{Bethe states}
\label{BS} \setcounter{equation}{0}

Up to now, the parameters $\a$ and $m$ in the definitions of the gauged operator $\overline{\mathscr{U}}(m,\a|u)$ in (\ref{def U left})
and the associated $K$-matrix $\overline{K}^+(m,\a|u)$ in (\ref{def K+ left}) (resp. $\mathscr{U}(m,\a|u)$ in (\ref{def U+ right})
and the associated $K$-matrix $K^-(m,\a|u)$ in (\ref{def K- right})) are arbitrary. The works in \cite{Bel13, Cao-14-Bethe-state,Bel14}
shed  light on the two important facts to construct the Bethe-type eigenstates of the $U(1)$-symmetry-broken integrable models:
(1) The inhomogeneous $T-Q$ relation plays a central role in constructing the
Bethe states because it enables one in this case to tell the wanted term  from the unwanted ones within the framework of the
algebraic Bethe Ansatz method;
(2) It also suggests that in order to construct the right Bethe states \footnote{Construction of left Bethe states is straightforward with a similar procedure.} of the transfer matrix (\ref{trans}),  one may choose the two
parameters $\a$ and $m$ according to the boundary parameters $\a_+$, $\b_+$ and $\theta_+$
to construct the generators (resp. according to the boundary parameters $\a_-$, $\b_-$ and $\theta_-$
to seek the associated reference state).

For this purpose, let us choose the gauge parameters in (\ref{def K+ left}) as follows
\bea
&&\quad\left\{
  \begin{array}{ll}
    \a\eta \stackrel{{\rm def}}{=}\a^{(l)}\eta=\eta-\theta_++i\frac{\pi}{2}\,\,{\rm mod}\,(2i\pi), \\[4pt]
    m\eta\stackrel{{\rm def}}{=}m^{(l)}\eta=\alpha_++\beta_+-i\frac{\pi}{2}\,\,{\rm mod}\,(2i\pi).
  \end{array}
\right.\label{case 1 K+}
\eea In this particular choice of the gauged parameters, the corresponding gauged $K$-matrix $\overline{K}^+(m,\a|u)$ given by
(\ref{def K+ left}) becomes diagonal
\bea
\overline{K}^+(m^{(l)},\a^{(l)}|u)={\rm Diag} (\overline{K}^+_{11}(m^{(l)},\a^{(l)}|u),\overline{K}^+_{22}(m^{(l)},\a^{(l)}|u)),\label{Diag-K+}
\eea where the non-vanishing matrix elements read
\begin{eqnarray}
&&\hspace{-1.4truecm}\overline{K}^+_{11}(m^{(l)},\a^{(l)}|u)\hspace{-0.08truecm}=\hspace{-0.08truecm}\frac{-2e^{-u}}{\cosh(\alpha_+
\hspace{-0.04truecm}+\hspace{-0.04truecm}\beta_+)}
\sinh(u\hspace{-0.04truecm}+\hspace{-0.04truecm}\alpha_+\hspace{-0.04truecm}+\hspace{-0.04truecm}\eta)
\cosh(u\hspace{-0.04truecm}+\hspace{-0.04truecm}\beta_+\hspace{-0.04truecm}+\hspace{-0.04truecm}\eta)
\cosh(\alpha_+\hspace{-0.04truecm}+\hspace{-0.04truecm}\beta_+\hspace{-0.04truecm}-\hspace{-0.04truecm}\eta),\\[2pt]
&&\hspace{-1.4truecm}\overline{K}^+_{22}(m^{(l)},\a^{(l)}|u)\hspace{-0.08truecm}=\hspace{-0.08truecm}
\frac{2e^{-u}}{\cosh(\alpha_+\hspace{-0.04truecm}+\hspace{-0.04truecm}\beta_+)}
\sinh(u\hspace{-0.04truecm}-\hspace{-0.04truecm}\alpha_+\hspace{-0.04truecm}+\hspace{-0.04truecm}\eta)
\cosh(u\hspace{-0.04truecm}-\hspace{-0.04truecm}\beta_+\hspace{-0.04truecm}+\hspace{-0.04truecm}\eta)
\cosh(\alpha_+\hspace{-0.04truecm}+\hspace{-0.04truecm}\beta_+\hspace{-0.04truecm}+\hspace{-0.04truecm}\eta).
\end{eqnarray} In this case the the transfer matrix (\ref{trans}) (see also (\ref{trans-1}) ) can  be rewritten as
\bea
t(u)&=&{\rm{tr}}\lt\{{K^+}(u)\mathscr{U}(u)\rt\}={\rm{tr}}\lt\{\overline{\mathscr{U}}(m^{(l)},\a^{(l)}|u)\overline{K}^+(m^{(l)},\a^{(l)}|u)\rt\}\no\\
&=&\overline{K}^+_{11}(m^{(l)},\a^{(l)}|u)\overline{\mathscr{A}}_{m^{(l)}}(u|\a^{(l)})
+\overline{K}^+_{22}(m^{(l)},\a^{(l)}|u)\overline{\mathscr{D}}_{m^{(l)}}(u|\a^{(l)}).\label{Simple-trans}
\eea
Direct calculation shows that the following identity holds
\begin{eqnarray}
&&{\overline{K}^+_{22}}(m^{(l)},\a^{(l)}|u)+\frac{\sinh\eta\sinh((m^{(l)}-1)\eta-2u)}{\sinh(2u+\eta)\sinh(m^{(l)}-1)\eta}
{\overline{K}^+_{11}}(m^{(l)},\a^{(l)}|u)\no\\[2pt]
&&\quad\quad\quad\quad=2e^{-u}\frac{\sinh(2u+2\eta)}{\sinh(2u+\eta)}\sinh(u-\alpha_+)\cosh(u-\beta_+).\label{K+ 11 22 right 1}
\end{eqnarray}
Then let us choose the gauge parameters in (\ref{def K- right}) such that the following relation is satisfied
\bea
(m^{(r)}+\a^{(r)})\eta=-\theta_-+\alpha_-+\beta_--N\eta+i\pi\,\,{\rm mod}\,(2i\pi).\label{case 1 K-}
\eea In this case the corresponding gauged $K$-matrix ${K}^-(m^{(r)}+N,\a^{(r)}|u)$ given by (\ref{def K- right}) becomes up-triangular with the
matrix element $K_{11}^-(m^{(r)}+N,\a^{(r)}|u)$ fixed, namely,
\bea
\hspace{-0.6truecm}K_{21}^-(m^{(r)}+N,\a^{(r)}|u)=0,\quad K_{11}^-(m^{(r)}+N,\a^{(r)}|u)=-2e^{u}\sinh(u-\alpha_-)\cosh(u-\beta_-).\label{Off-diagonal-K-}
\eea
Although neither the parameter $\a^{(r)}$ nor $m^{(r)}$ is fixed by the up-triangularity condition of ${K}^-(m^{(r)},\a^{(r)}|u)$,
the sum of the two parameters is unique as shown in (\ref{case 1 K-}). This allows us to define a unique reference state $|\Omega\rangle$,
\bea
|\Omega\rangle=|\a^{(r)}+m^{(r)}\rangle, \label{Reference-state}
\eea
where the  state $|\a^{(r)}+m^{(r)}\rangle$ is defined by (\ref{Right-vacumm}) with the parameter $\a+m$ fixed by the boundary parameters (see (\ref{case 1 K-})).
It should be noted that the reference state $|\Omega\rangle$ is rather different from that used in algebraic Bethe ansatz  (namely, the all spin-up or spin-down state \cite{Skl88, Kor93}).

Let  $|\Psi\rangle$ be an eigenstate of the transfer matrix $t(u)$ with an eigenvalue $\L(u)$, namely,
\bea
t(u)\,|\Psi\rangle=\L(u)\, |\Psi\rangle.
\eea
Due to the fact that the left states $\{\langle\a^{(l)},m^{(l)};\theta_{p_1},\cdots,\theta_{p_n}||n=0,\cdots, N,\,
1\leq p_1<p_2<\cdots<p_n\leq N\}$ given by (\ref{def state left}) form a basis of the dual Hilbert
space, the eigenstate $|\Psi\rangle$ is completely determined (up to an overall scalar factor) by the following scalar products \cite{Cao1,Cao-14-Bethe-state}
\begin{eqnarray}
F_n(\theta_{p_1},\cdots,\theta_{p_n})={\langle}\a^{(l)},m^{(l)};\theta_{p_1},\cdots,\theta_{p_n}|\Psi\rangle,\quad n=0,\cdots,N.
\end{eqnarray}
After a tedious calculation, we have that the above scalar products are given by
\bea
\hspace{-1.2truecm}F_n(\theta_{p_1},\cdots,\theta_{p_n})\hspace{-0.12truecm}&=&\hspace{-0.12truecm}
\prod_{j=1}^n\hspace{-0.04truecm}\lt\{\frac{\hspace{-0.04truecm}-\sinh(2\theta_{p_j}-\eta)\Lambda(-\theta_{p_j})e^{-\theta_{p_j}}}
{2\sinh(2\theta_{p_j}-2\eta)\sinh(\theta_{p_j}\hspace{-0.04truecm}+\hspace{-0.04truecm}\a_+)
\cosh(\theta_{p_j}\hspace{-0.04truecm}+\hspace{-0.04truecm}\b_+)}\rt\}F_0,\no\\
n&=&0,1,\cdots,N,\label{Psi-1}
\eea where $ F_0=\langle\a^{(l)},m^{(l)}|\Psi\rangle$ is an overall scalar factor. The proof of the above relations is relegated to Appendix C.

Following the method developed in \cite{Cao-14-Bethe-state}, we propose that the Bethe-type eigenstate of the transfer matrix (\ref{trans}) for the
present model is given by
\bea
|\l_1,\cdots,\l_N\rangle=\overline{\mathscr{C}}_{m^{(l)}}(\l_1|\a^{(l)})\, \overline{\mathscr{C}}_{m^{(l)}+2}(\l_2|\a^{(l)})\cdots
\overline{\mathscr{C}}_{m^{(l)}+2(N-1)}(\l_N|\a^{(l)})\,|\Omega\rangle,\label{Bethe-right}
\eea where the two parameters $\a^{(l)}$ and $m^{(l)}$ are given by (\ref{case 1 K+}) and the $N$ parameters $\{\l_j|j=1,\cdots,N\}$ satisfy the BAEs (\ref{BAE-1}).
We shall show that the chosen reference state $|\Omega\rangle$ given by (\ref{Reference-state}) indeed makes the conditions (\ref{Psi-1}) fulfilled.
For an eigenvalue $\Lambda(u)$ given by
the inhomogeneous $T-Q$ relation (\ref{T-Q-1}), its value at the point $-\theta_{j}$ takes a simple form:
\bea
\Lambda(-\theta_j)=a(-\theta_j)\frac{Q(-\theta_j-\eta)}{Q(-\theta_j)},\quad j=1,\cdots,N.
\eea
The above relations and the equations (\ref{eigenvalue C left}) imply that the conditions (\ref{Psi-1}) are equivalent to the following
requirements on the reference state:
\bea
{\langle}\a^{(l)},m';\theta_{p_1},\cdots,\theta_{p_{n}}|\Omega\rangle&=&\hspace{-0.4truecm}\prod_{j=1}^n
\lt\{2e^{-\theta_{p_{j}}}\sinh(\theta_{p_{j}}+\alpha_-)\cosh(\theta_{p_{j}}+\beta_-)\bar A(-\theta_{p_{j}})\rt\}
\frac{F_0}{G_0},\no\\
n&=&0,1,\cdots,N,\label{Reference-state-1}
\eea  where $m'=m^{(l)}+2N$ and the overall coefficient $G_0$ independent upon $n$ is
\bea
G_0=\prod_{j=1}^Ng_0(\l_j|m^{(l)}+2(j-1),\a^{(l)}),\label{G-0}
\eea
with function $g_0(u|m,\a)$ given by (\ref{g-0}).  Actually, the above conditions uniquely determine the reference state $|\Omega\rangle$ up to a scalar factor. Direct calculation shows that the
state $|\Omega\rangle$  given by (\ref{Reference-state}) indeed satisfies the conditions (\ref{Reference-state-1}). The proof  is relegated to Appendix C.
Finally, we conclude that the Bethe state $|\l_1,\cdots,\l_N\rangle$ becomes an eigenstate of the transfer matrix $t(u)$ with the
eigenvalue $\L(u)$ given by (\ref{T-Q-1}) provided that the reference state $|\Omega\rangle$ is given by (\ref{Reference-state}) and  the $N$ parameters $\{\l_j|j=1,\cdots,N\}$ satisfy the BAEs (\ref{BAE-1}).

From the definitions (\ref{Intertwiner-1})-(\ref{Intertwiner-2}) of the gauge matrices, it is clear that both the reference state $|\Omega\rangle$ and the generators $\overline{\mathscr{C}}_{m^{(l)}+2j}(u|\a^{(l)})$ have well-defined homogeneous limits: $\{\theta_j\rightarrow 0\}$.  This implies that
the homogeneous limit of the Bethe state (\ref{Bethe-right}) exactly gives rise to the corresponding Bethe state of the homogeneous  XXZ
spin-$\frac{1}{2}$ chain with arbitrary boundary fields, where the associated $T-Q$ relation and BAEs are given by (\ref{T-Q-1}) and (\ref{BAE-1}) with $\{\theta_j=0\}$.
It would be interesting to study the relation between our Bethe states and the eigenstates proposed in \cite{Fad14} for which the homogeneous limit is still unclear.

%%%%%%%%%%%%%%%%%%%%%%%%%%%%%%%%%%%%%%%%%%%%%%%%%%%%%%%%%%%%%%%
%                                                             %
%  7. Conclusions                                             %
%                                                             %
%                                                             %
%                                                             %
%%%%%%%%%%%%%%%%%%%%%%%%%%%%%%%%%%%%%%%%%%%%%%%%%%%%%%%%%%%%%%%

\section{Conclusions}
\label{Con} \setcounter{equation}{0}

It should be emphasized
that constructing the Bethe state of $U(1)$-symmetry-broken models  had challenged for many
years because of the lacking of the inhomogeneous $T-Q$ relations such as (\ref{T-Q-1}).
The idea of this paper to construct the Bethe state is to search for two gauge transformations such that one makes
the resulting $K^+$-matrix to be diagonal  and the other makes the resulting $K^-$-matrix up-triangular. Then we find that
the two parameters $m^{(l)}$ and $\a^{(l)}$ of the first gauge transformation must obey the following equations
\begin{eqnarray}
\lt\{\begin{array}{l}\sinh(\alpha_++\beta_+)=\sinh(\theta_++(\alpha^{(l)}-1)\eta+m^{(l)}\eta),\\[2pt]
\sinh(\alpha_++\beta_+)=\sinh(\theta_++(\alpha^{(l)}-1)\eta-m^{(l)}\eta),\\
\end{array}\rt.\label{condition K+ left}
\end{eqnarray} while the parameters of the second gauge transformation have to satisfy the relation
\bea
\sinh(\alpha_-+\beta_-)+\sinh(\theta_-+(m^{(r)}+\alpha^{(r)})\eta+N\eta)=0.\label{condition K+ right}
\eea The equation (\ref{condition K+ left}) is to determine the generators $\overline{\mathscr{C}}_{m^{(l)}+2j}(u|\a^{(l)})$,
while the equation (\ref{condition K+ right}) is to choose the associated reference state (such as (\ref{Reference-state})). It is found
that besides the solution given by (\ref{case 1 K+}) and (\ref{case 1 K-})  there exist three other  solutions of (\ref{condition K+ left})
and (\ref{condition K+ right}). Each of the three solutions gives rise to a set of Bethe states with eigenvalues parameterized by a
$T-Q$ relation of the form (\ref{T-Q-1}) by replacing
$\a_{\pm}$, $\b_{\pm}$ with $\pm\a_{\pm}$, $\pm\b_{\pm}$. Nevertheless, different types of inhomogeneous $T-Q$  relations \cite{Yan06, Cao2} only give different parameterizations of the eigenvalues of the transfer matrix but not new solutions. We note that for the degenerate case considered in \cite{Cao03}, the present method may not work but the Bethe states can be obtained via generalized algebraic Bethe Ansatz.

%%%%%%%%%%%%%%%%%%%%%%%%%%%%%%%%%%%%%%%%%%%%%%%%%%%%%%%%%%%%%%%
%                                                             %
%  Acknowledgments                                            %
%                                                             %
%%%%%%%%%%%%%%%%%%%%%%%%%%%%%%%%%%%%%%%%%%%%%%%%%%%%%%%%%%%%%%%
\section*{Acknowledgments}

The financial supports from the National Natural Science Foundation
of China (Grant Nos. 11375141, 11374334, 11434013, 11425522), the
National Program for Basic Research of MOST (973 project under grant
No.2011CB921700), BCMIIS and the Strategic Priority Research Program
of the Chinese Academy of Sciences are gratefully acknowledged. Two
of the authors (W.-L. Yang and K. Shi) would like to thank IoP/CAS
for the hospitality.

%%%%%%%%%%%%%%%%%%%%%%%%%%%%%%%%%%%%%%%%%%%%%%%%%%%%%%%%%%%%%%%%
%                                                             %
%  Appendix A                                                 %
%                                                             %
%                                                             %
%                                                             %
%%%%%%%%%%%%%%%%%%%%%%%%%%%%%%%%%%%%%%%%%%%%%%%%%%%%%%%%%%%%%%%

\section*{Appendix A: Intertwining relations  }
\setcounter{equation}{0}
\renewcommand{\theequation}{A.\arabic{equation}}

We list some intertwining relations (or face-vertex correspondence relations in \cite{Cao03}) which are useful to construct the reference state and the commutation relations among the
gauged operators:\footnote{In fact these vectors depend also on $\a$ but as this parameter will not vary in the following
relations, in this appendix we omit this argument for simplicity temporarily.}
\begin{eqnarray}
&&\hspace{-1.8truecm}R_{12}(u_1-u_2)X^1_{m+2}(u_1)X^2_{m+1}(u_2)=\frac{\sinh(u_1-u_2+\eta)}{\sinh\eta}X^2_{m+2}(u_2)X^1_{m+1}(u_1),\label{relation1}\\
&&\hspace{-1.8truecm}R_{12}(u_1-u_2)X^1_{m}(u_1)Y^2_{m-1}(u_2)=\frac{\sinh(u_1-u_2)\sinh(m-1)\eta}{\sinh\eta\sinh m\eta}Y^2_m(u_2)X^1_{m+1}(u_1)\no\\
&&\qquad\qquad\qquad\qquad\qquad\qquad\quad+\frac{\sinh(m\eta+u_1-u_2)}{\sinh m\eta}X^2_m(u_2)Y^1_{m-1}(u_1),\label{relation2}\\[2pt]
&&\hspace{-1.8truecm}R_{12}(u_1-u_2)Y^1_{m}(u_1)X^2_{m+1}(u_2)=\frac{\sinh(u_1-u_2)\sinh(m+1)\eta}{\sinh\eta\sinh m\eta}X^2_m(u_2)Y^1_{m-1}(u_1)\no\\
&&\qquad\qquad\qquad\qquad\qquad\qquad\quad+\frac{\sinh(m\eta-u_1+u_2)}{\sinh m\eta}Y^2_m(u_2)X^1_{m+1}(u_1),\label{com relation3}\\[2pt]
&&\hspace{-1.8truecm}R_{12}(u_1-u_2)Y^1_{m-2}(u_1)Y^2_{m-1}(u_2)=\frac{\sinh(u_1-u_2+\eta)}{\sinh\eta}Y^2_{m-2}(u_2)Y^1_{m-1}(u_1),\label{relation4}\\
&&\hspace{-1.8truecm}R_{12}(u_1-u_2)\widehat{X}^2_{m-1}(u_2)\widehat{X}^1_{m}(u_1)=\frac{\sinh(u_1-u_2+\eta)}{\sinh\eta}\widehat{X}_{m}^2(u_2)\widehat{X}_{m-1}^1(u_1),
\label{relations5}\\[2pt]
&&\hspace{-1.8truecm}R_{12}(u_1-u_2)\widehat{X}^2_{m-1}(u_2)\widehat{Y}^1_{m+2}(u_1)=\frac{\sinh(u_1-u_2)\sinh(m+1)\eta}{\sinh\eta\sinh m\eta}\widehat{X}_{m-2}^2(u_2)\widehat{Y}^1_{m+1}(u_1)\no\\
&&\qquad\qquad\qquad\qquad\qquad\qquad\quad+
\frac{\sinh(m\eta-u_1+u_2)}{\sinh m\eta}\widehat{Y}_{m+2}^2(u_2)\widehat{X}_{m-1}^1(u_1),\label{relation6}\\[2pt]
&&\hspace{-1.8truecm}R_{12}(u_1-u_2)\widehat{Y}^2_{m+1}(u_2)\widehat{X}^1_{m-2}(u_1)=\frac{\sinh(u_1-u_2)\sinh(m-1)\eta}{\sinh\eta\sinh m\eta}\widehat{Y}_{m+2}^2(u_2)\widehat{X}_{m-1}^1(u_1)\no\\
&&\qquad\qquad\qquad\qquad\qquad\qquad\quad+\frac{\sinh(m\eta+u_1-u_2)}{\sinh m\eta}\widehat{X}_{m-2}^2(u_2)\widehat{Y}^1_{m+1}(u_1),\label{relation7}\\[2pt]
&&\hspace{-1.8truecm}R_{12}(u_1-u_2)\widehat{Y}^2_{m+1}(u_2)\widehat{Y}^1_{m}(u_1)=
\frac{\sinh(u_1-u_2+\eta)}{\sinh\eta}\widehat{Y}_m^2(u_2)\widehat{Y}^1_{m+1}(u_1),\label{relation8}
\end{eqnarray}
\begin{eqnarray}
&&\hspace{-1.2truecm}\overline{X}^1_{m-1}(u_1)\overline{X}^2_{m-2}(u_2)R_{12}(u_1-u_2)=
\frac{\sinh(u_1-u_2+\eta)}{\sinh\eta}\overline{X}^2_{m-1}(u_2)\overline{X}_{m-2}^1(u_1),\label{relation9}\\[2pt]
&&\hspace{-1.2truecm}\overline{X}^1_{m-1}(u_1)\overline{Y}^2_{m}(u_2)R_{12}(u_1-u_2)=
\frac{\sinh(u_1-u_2)\sinh(m+1)\eta}{\sinh\eta\sinh m\eta}\overline{Y}_{m+1}^2(u_2)\overline{X}^1_m(u_1)\no\\
&&\qquad\qquad\qquad\qquad\qquad\qquad\quad\quad+\frac{\sinh(m\eta+u_1-u_2)}{\sinh m\eta}\overline{X}_{m-1}^2(u_2)\overline{Y}_m^1(u_1),\label{relations10}\\[2pt]
&&\hspace{-1.2truecm}\overline{Y}^1_{m+1}(u_1)\overline{X}^2_m(u_2)R_{12}(u_1-u_2)=
\frac{\sinh(u_1-u_2)\sinh(m-1)\eta}{\sinh\eta\sinh m\eta}\overline{X}^2_{m-1}(u_2)\overline{Y}^1_m(u_1)\no\\
&&\qquad\qquad\qquad\qquad\qquad\qquad\quad\quad+\frac{\sinh(m\eta-u_1+u_2)}{\sinh m\eta}\overline{Y}_{m+1}^2(u_2)\overline{X}^1_m(u_1),\label{relation11}\\[2pt]
&&\hspace{-1.2truecm}\overline{Y}^1_{m+1}(u_1)\overline{Y}^1_{m+2}(u_2)R_{12}(u_1-u_2)=
\frac{\sinh(u_1-u_2+\eta)}{\sinh\eta}\overline{Y}^2_{m+1}(u_2)\overline{Y}^1_{m+2}(u_1),\label{relations12}\\[2pt]
&&\hspace{-1.2truecm}\widetilde{X}^1_{m+1}(u_1)\widetilde{X}^2_m(u_2)R_{12}(u_1-u_2)=
\frac{\sinh(u_1-u_2+\eta)}{\sinh\eta}\widetilde{X}^2_{m+1}(u_2)\widetilde{X}^1_m(u_1),\label{relation13}\\[2pt]
&&\hspace{-1.2truecm}\widetilde{X}^1_{m+1}(u_1)\widetilde{Y}^2_{m-2}(u_2)R_{12}(u_1-u_2)
=\frac{\sinh(u_1-u_2)\sinh(m+1)\eta}{\sinh\eta\sinh m\eta}\widetilde{Y}^2_{m-1}(u_2)\widetilde{X}^1_{m+2}(u_1)\no\\
&&\qquad\qquad\qquad\qquad\qquad\qquad\quad\quad+
\frac{\sinh(m\eta+u_1-u_2)}{\sinh m\eta}\widetilde{X}^2_{m+1}(u_2)\widetilde{Y}^1_{m-2}(u_1),\label{realtion14}\\[2pt]
&&\hspace{-1.2truecm}\widetilde{Y}^1_{m-1}(u_1)\widetilde{X}^2_{m+2}(u_2)R_{12}(u_1-u_2)
=\frac{\sinh(u_1-u_2)\sinh(m-1)\eta}{\sinh\eta\sinh m\eta}\widetilde{X}^2_{m+1}(u_2)\widetilde{Y}^1_{m-2}(u_1)\no\\
&&\qquad\qquad\qquad\qquad\qquad\qquad\quad\quad+
\frac{\sinh(m\eta-u_1+u_2)}{\sinh m\eta}\widetilde{Y}^2_{m-1}(u_2)\widetilde{X}^1_{m+2}(u_1),\label{relations15}\\[2pt]
&&\hspace{-1.2truecm}\widetilde{Y}^1_{m-1}(u_1)\widetilde{Y}^2_{m}(u_2)R_{12}(u_1-u_2)=\frac{\sinh(u_1-u_2+\eta)}
{\sinh\eta}\widetilde{Y}^2_{m-1}(u_2)\widetilde{Y}^1_m(u_1),\label{relations16}
\end{eqnarray}
\begin{eqnarray}
&&\hspace{-1.2truecm}\overline{X}^2_m(u_2)R_{12}(u_1\hspace{-0.04truecm}-\hspace{-0.04truecm}u_2)X_m^1(u_1)=
\frac{\sinh(u_1\hspace{-0.04truecm}-\hspace{-0.04truecm}u_2)
\sinh(m\hspace{-0.04truecm}-\hspace{-0.04truecm}1)\eta}{\sinh\eta\sinh m\eta}
\overline{X}^2_{m\hspace{-0.04truecm}-\hspace{-0.04truecm}1}(u_2)
X^1_{m\hspace{-0.04truecm}+\hspace{-0.04truecm}1}(u_1),\label{relation17}\\[2pt]
&&\hspace{-1.2truecm}\overline{X}^2_m(u_2)R_{12}(u_1-u_2)Y_m^1(u_1)
=\frac{\sinh(u_1-u_2+\eta)}{\sinh\eta}\overline{X}^2_{m+1}(u_2)Y^1_{m+1}(u_1)\no\\
&&\qquad\qquad\qquad\qquad\qquad\qquad\quad\quad+
\frac{\sinh(m\eta-u_1+u_2)}{\sinh m\eta}\overline{Y}^2_{m+1}(u_2)X^1_{m+1}(u_1),\label{relation18}\\[2pt]
&&\hspace{-1.2truecm}\overline{Y}^2_m(u_2)R_{12}(u_1-u_2)X^1_m(u_1)
=\frac{\sinh(u_1-u_2+\eta)}{\sinh\eta}\overline{Y}^2_{m-1}(u_2)X^1_{m-1}(u_1)\no\\
&&\qquad\qquad\qquad\qquad\qquad\qquad\quad\quad+
\frac{\sinh(m\eta+u_1-u_2)}{\sinh m\eta}\overline{X}^2_{m-1}(u_2)Y^1_{m-1}(u_1),\label{relation19}\\[2pt]
&&\hspace{-1.2truecm}\overline{Y}^2_m(u_2)R_{12}(u_1-u_2)Y^1_m(u_1)=
\frac{\sinh(u_1-u_2)\sinh(m+1)\eta}{\sinh\eta\sinh m\eta}\overline{Y}^2_{m+1}(u_2)Y^1_{m-1}(u_1),\label{relation20}\\[2pt]
&&\hspace{-1.2truecm}\widetilde{X}^1_{m\hspace{-0.04truecm}+\hspace{-0.04truecm}1}(u_1)
R_{12}(u_1\hspace{-0.04truecm}-\hspace{-0.04truecm}u_2)X^2_{m\hspace{-0.04truecm}+\hspace{-0.04truecm}1}(u_2)=\hspace{-0.08truecm}
\frac{\sinh(u_1\hspace{-0.04truecm}-\hspace{-0.04truecm}u_2)
\sinh(m\hspace{-0.04truecm}+\hspace{-0.04truecm}1)\eta}
{\sinh\eta\sinh m\eta}X^2_m(u_2)\widetilde{X}^1_{m\hspace{-0.04truecm}+\hspace{-0.04truecm}2}(u_1),
\label{relation21}\\[2pt]
&&\hspace{-1.2truecm}\widetilde{X}^1_{m+1}(u_1)R_{12}(u_1-u_2)Y^2_{m-1}(u_2)
=\frac{\sinh(u_1-u_2+\eta)}{\sinh\eta}Y^2_{m-2}(u_2)\widetilde{X}^1_m(u_1)\no\\
&&\qquad\qquad\qquad\qquad\qquad\qquad\quad\quad+
\frac{\sinh(m\eta+u_1-u_2)}{\sinh m\eta}X^2_m(u_2)\widetilde{Y}^1_{m-2}(u_1),\label{relation22}\\[2pt]
&&\hspace{-1.2truecm}\widetilde{Y}^1_{m-1}(u_1)R_{12}(u_1-u_2)X^2_{m+1}(u_2)
=\frac{\sinh(u_1-u_2+\eta)}{\sinh\eta}X^2_{m+2}(u_2)\widetilde{Y}^1_m(u_1)\no\\
&&\qquad\qquad\qquad\qquad\qquad\qquad\quad\quad+
\frac{\sinh(m\eta-u_1+u_2)}{\sinh m\eta}Y^2_m(u_2)\widetilde{X}^1_{m+2}(u_1),\label{relation23}\\[2pt]
&&\hspace{-1.2truecm}\widetilde{Y}^1_{m\hspace{-0.04truecm}-\hspace{-0.04truecm}1}(u_1)
R_{12}(u_1\hspace{-0.04truecm}-\hspace{-0.04truecm}u_2)
Y^2_{m\hspace{-0.04truecm}-\hspace{-0.04truecm}1}(u_2)=
\frac{\sinh(u_1\hspace{-0.04truecm}-\hspace{-0.04truecm}u_2)
\sinh(m\hspace{-0.04truecm}-\hspace{-0.04truecm}1)\eta}{\sinh\eta\sinh m\eta}Y^2_m(u_2)
\widetilde{Y}^1_{m\hspace{-0.04truecm}-\hspace{-0.04truecm}2}(u_1),\label{realtion24}\\[2pt]
&&\hspace{-1.2truecm}\overline{X}^1_{m\hspace{-0.04truecm}-\hspace{-0.04truecm}1}(u_1)
R_{12}(u_1\hspace{-0.04truecm}-\hspace{-0.04truecm}u_2)
\widehat{X}^2_{m\hspace{-0.04truecm}-\hspace{-0.04truecm}1}(u_2)=\hspace{-0.04truecm}
\frac{\sinh(u_1\hspace{-0.04truecm}-\hspace{-0.04truecm}u_2)
\sinh(m\hspace{-0.04truecm}+\hspace{-0.04truecm}1)\eta}{\sinh\eta\sinh m\eta}
\widehat{X}_{m\hspace{-0.04truecm}-\hspace{-0.04truecm}2}^2(u_2)\overline{X}_m^1(u_1),\label{relation25}\\[2pt]
&&\hspace{-1.2truecm}\overline{X}^1_{m-1}(u_1)R_{12}(u_1-u_2)\widehat{Y}^2_{m+1}(u_2)
=\frac{\sinh(u_1-u_2+\eta)}{\sinh\eta}\widehat{Y}_m^2(u_2)\overline{X}_{m-2}^1(u_1)\no\\
&&\qquad\qquad\qquad\qquad\qquad\qquad\quad\quad+
\frac{\sinh(m\eta+u_1-u_2)}{\sinh m\eta}\widehat{X}_{m-2}^2(u_2)\overline{Y}_m^1(u_1),\label{relation26}\\[2pt]
&&\hspace{-1.2truecm}\overline{Y}^1_{m+1}(u_1)R_{12}(u_1-u_2)\widehat{X}_{m-1}^2(u_2)
=\frac{\sinh(u_1-u_2+\eta)}{\sinh\eta}\widehat{X}_{m}^2(u_2)\overline{Y}_{m+2}^1(u_1)\no\\
&&\qquad\qquad\qquad\qquad\qquad\qquad\quad\quad+
\frac{\sinh(m\eta-u_1+u_2)}{\sinh m\eta}\widehat{Y}_{m+2}^2(u_2)\overline{X}_m^1(u_1),\label{relation27}\\[2pt]
&&\hspace{-1.2truecm}\overline{Y}^1_{m\hspace{-0.04truecm}+\hspace{-0.04truecm}1}(u_1)
R_{12}(u_1\hspace{-0.04truecm}-\hspace{-0.04truecm}u_2)
\widehat{Y}_{m\hspace{-0.04truecm}+\hspace{-0.04truecm}1}^2(u_2)
\hspace{-0.04truecm}=\hspace{-0.04truecm}\frac{\sinh(u_1\hspace{-0.04truecm}-\hspace{-0.04truecm}u_2)
\sinh(m\hspace{-0.04truecm}-\hspace{-0.04truecm}1)\eta}{\sinh\eta\sinh m\eta}
\widehat{Y}_{m\hspace{-0.04truecm}+\hspace{-0.04truecm}2}^2(u_2)\overline{Y}_m^1(u_1),\label{relation28}
\end{eqnarray}
where $X^1_m(u)$, $X^2_m(u)$ are embedding vectors in  the $1$-st and $2$-nd tensor space, respectively. Moreover, the vectors
also enjoy the following  orthonormal relations:
\begin{eqnarray}
&&\overline{Y}_m(u)X_m(u)=1,\quad\quad \overline{Y}_m(u)Y_m(u)=0,\no\\[2pt]
&&\overline{X}_m(u)X_m(u)=0,\quad\quad \overline{X}_m(u)Y_m(u)=1,\no\\[2pt]
&&X_{m}(u)\overline{Y}_m(u)+Y_{m}(u)\overline{X}_m(u)=\left(
                                                      \begin{array}{cc}
                                                        1 & 0 \\
                                                        0 & 1 \\
                                                      \end{array}
                                                    \right),\label{orth M}\\[4pt]
&&\widetilde{Y}_{m-1}(u)X_{m+1}(u)=1,\quad\quad \widetilde{Y}_{m-1}(u)Y_{m-1}(u)=0,\no\\[2pt]
&&\widetilde{X}_{m+1}(u)X_{m+1}(u)=0,\quad\quad \widetilde{X}_{m+1}(u)Y_{m-1}(u)=1,\no\\[2pt]
&&X_{m+1}(u)\widetilde{Y}_{m-1}(u)+Y_{m-1}(u)\widetilde{X}_{m+1}(u)=\left(
                                                      \begin{array}{cc}
                                                        1 & 0 \\
                                                        0 & 1 \\
                                                      \end{array}
                                                    \right),\label{orth tilde M}\\[4pt]
&&\overline{Y}_{m+1}(u)\widehat{X}_{m-1}(u)=1, \quad\quad\overline{Y}_{m+1}(u)\widehat{Y}_{m+1}(u)=0,\no\\[2pt]
&&\overline{X}_{m-1}(u)\widehat{X}_{m-1}(u)=0, \quad\quad\overline{X}_{m-1}(u)\widehat{Y}_{m+1}(u)=1,\no\\[2pt]
&&\widehat{X}_{m-1}(u)\overline{Y}_{m+1}(u)+\widehat{Y}_{m+1}(u)\overline{X}_{m-1}(u)=\left(
                                                          \begin{array}{cc}
                                                            1 & 0 \\
                                                            0 & 1 \\
                                                          \end{array}
                                                        \right).\label{orth hat M}
\end{eqnarray}

%%%%%%%%%%%%%%%%%%%%%%%%%%%%%%%%%%%%%%%%%%%%%%%%%%%%%%%%%%%%%%%%
%                                                             %
%  Appendix B                                                 %
%                                                             %
%                                                             %
%                                                             %
%%%%%%%%%%%%%%%%%%%%%%%%%%%%%%%%%%%%%%%%%%%%%%%%%%%%%%%%%%%%%%%

\section*{Appendix B: Commutation relations }
\setcounter{equation}{0}
\renewcommand{\theequation}{B.\arabic{equation}}

Using QYBE (\ref{QYB}) and the RE (\ref{RE-V}), one may derive that
\begin{eqnarray}
R_{12}(u_1-u_2)\mathscr{U}_1(u_1)R_{21}(u_1+u_2)\mathscr{U}_2(u_2)=\mathscr{U}_2(u_2)R_{21}(u_1+u_2)\mathscr{U}_1(u_1)R_{12}(u_1-u_2).\label{U relation}
\end{eqnarray}
Multiplying the above equation with $\overline{X}^1_{m+1}(u_1)\overline{X}_m^2(u_2)$ from the left and $\widehat{X}^1_{m+1}(-u_1)\widehat{X}^2_{m+2}(-u_2)$ from the right, and using the relations (\ref{relations5}) and (\ref{relation9}), we arrive at (\ref{CC relation left}). Similarly,
multiplying (\ref{U relation}) with $\overline{X}^1_{m-1}(u_1)\overline{X}_{m-2}^2(u_2)$ ($\overline{X}^1_{m-1}(u_1)\overline{X}_{m-2}^2(u_2)$) from the left and $\widehat{Y}^1_{m+1}(-u_1)\widehat{Y}^2_{m}(-u_2)$ ($\widehat{X}^1_{m-1}(-u_1)\widehat{Y}^2_{m+2}(-u_2)$ )from the right and using the intertwining relations (\ref{relation1})-(\ref{relation28}),
one can obtain the relation (\ref{DD relation left}) (or (\ref{DC relation left})). Using the similar method and the relation (\ref{DD relation left}),
one can further  check (\ref{DA relation 2}).

%%%%%%%%%%%%%%%%%%%%%%%%%%%%%%%%%%%%%%%%%%%%%%%%%%%%%%%%%%%%%%%%
%                                                             %
%  Appendix C                                                 %
%                                                             %
%                                                             %
%                                                             %
%%%%%%%%%%%%%%%%%%%%%%%%%%%%%%%%%%%%%%%%%%%%%%%%%%%%%%%%%%%%%%%

\section*{Appendix C: Proof the Bethe state }
\setcounter{equation}{0}
\renewcommand{\theequation}{C.\arabic{equation}}

There are several ways \cite{Bel13, Cao-14-Bethe-state, Bel14, Cra14} to show that the state $|\l_1,\cdots,\l_N\rangle$ constructed
by (\ref{Bethe-right}) is an eigenstate of the transfer matrix (\ref{trans}). Here we adopt the method developed in \cite{Cao-14-Bethe-state}
to demonstrate it.
\subsection*{C.1 The proof of (\ref{Psi-1})}
For arbitrary parameters $\a,\,m$ let us introduce the following  left states\footnote{Such states were used as a basis to construct the  SoV eigenstates of the XXZ open chain \cite{Fad14}. Here we use two different gauge transformations respectively for the left and right reference states to reach the Bethe states.}
parameterized by the $N$ inhomogeneous parameters $\{\theta_j\}$:
\begin{eqnarray}
\langle \a,m; \theta_{p_1}\cdots\theta_{p_n}|&=&\langle{\a+m}|\overline{\mathscr{D}}_{m}(-\theta_{p_1}|\a)
\cdots\overline{\mathscr{D}}_{m}(-\theta_{p_n}|\a),\no\\
 1\leq q_1&<&q_2<\ldots<q_n\leq N,\quad n=0,1,\cdots,N.\label{def state left}
\end{eqnarray}
The commutation relations (\ref{DD relation left}), (\ref{DC relation left}) and (\ref{C left function}) imply that
\begin{eqnarray}
\langle\a,m;\theta_{p_1},\cdots,\theta_{p_n}|\,\overline{\mathscr{C}}_{m}(u|\a)
\hspace{-0.08truecm}=\hspace{-0.08truecm}g(u,\{\theta_{p_1},\cdots,\theta_{p_n}\})
{\langle}\a,m\hspace{-0.08truecm}+\hspace{-0.08truecm}2;\theta_{p_1},\cdots,\theta_{p_n}|,
\label{eigenvalue C left}
\end{eqnarray}
where
\begin{eqnarray}
g(u,\{\theta_{p_1},\cdots,\theta_{p_n}\})\hspace{-0.4truecm}&=&g_0(u|m,\a)\,
\prod_{j=1}^n\frac{\sinh(u+\theta_{p_j}+\eta)\sinh(u-\theta_{p_j})}{\sinh(u-\theta_{p_j}+\eta)\sinh(u+\theta_{p_j})},\label{def bar g}
\end{eqnarray}
and
\bea
g_0(u|m,\a)&=&{\overline{K}^-_{21}}
(m\hspace{-0.04truecm}-\hspace{-0.04truecm}N;\hspace{-0.04truecm}\a|u)
\frac{\sinh(m\hspace{-0.04truecm}+\hspace{-0.04truecm}2)\eta}
{\sinh(m\hspace{-0.04truecm}+\hspace{-0.04truecm}2\hspace{-0.04truecm}-\hspace{-0.04truecm}N)\eta}\no\\
&&\times\prod_{j=1}^N\frac{\sinh(u\hspace{-0.04truecm}-\hspace{-0.04truecm}\theta_j\hspace{-0.04truecm}+\hspace{-0.04truecm}\eta)
\sinh(u\hspace{-0.04truecm}+\hspace{-0.04truecm}\theta_j)}{\sinh^2\eta}. \label{g-0}
\eea
The above equations also lead to the following fact
\bea
\langle\a,m;\theta_{p_1},\cdots,\theta_{p_n}|\,\overline{\mathscr{C}}_{m}(-\theta_{p_{j}}|\a)=0,
\quad j\neq 1,\cdots,n.\label{Vanishing-cond-1}
\eea

Keeping  the particular choice of
the parameters  (\ref{case 1 K+}) and the simple decomposition (\ref{Simple-trans}) of the transfer matrix, one can derive the following recursive relations (see (\ref{Recursive-relation-1}) below) by considering
the quantity of $\langle \a^{(l)},m^{(l)};\theta_{p_1},\cdots,\theta_{p_n}|t(-\theta_{p_{n+1}})|\Psi\rangle$,
\begin{eqnarray}
&&\hspace{-0.8truecm}\Lambda(-\theta_{p_{n+1}})F_n(\theta_{p_1},\cdots,\theta_{p_n})\no\\
&&\quad\quad=\overline{K}^+_{11}(m^{(l)},\a^{(l)}|-\theta_{p_{n+1}})
\langle \a^{(l)},m^{(l)};\theta_{p_1},\cdots,\theta_{p_n}|\mathscr{A}_{m^{(l)}}(-\theta_{p_{n+1}}|\a^{(l)})|\Psi\rangle\no\\[4pt]
&&\,\,\quad\quad\quad+\overline{K}^+_{22}(m^{(l)},\a^{(l)}|-\theta_{p_{n+1}})F_{n+1}(\theta_{p_1},\cdots,\theta_{p_n},\theta_{p_{n+1}}).\no
\end{eqnarray}
The relations (\ref{DC relation left}), (\ref{DA relation 2}), (\ref{DA relation special}) and (\ref{Vanishing-cond-1}) enable us to
further simplify the above equation
\bea
\Lambda(-\theta_{p_{n+1}})F_n(\theta_{p_1},\cdots,\theta_{p_n})\hspace{-0.4truecm}&=&\hspace{-0.4truecm}
F_{n+1}(\theta_{p_1},\cdots,\theta_{p_n},\theta_{p_{n+1}})\lt\{\overline{K}^+_{22}(m^{(l)},\a^{(l)}|-\theta_{p_{n+1}})\rt.
\no\\[4pt]
&&-\lt.\frac{\sinh((m^{(l)}\hspace{-0.04truecm}-\hspace{-0.04truecm}1)\eta\hspace{-0.04truecm}+\hspace{-0.04truecm}2\theta_{p_{n+1}})\sinh\eta}
{\sinh(m^{(l)}\hspace{-0.04truecm}-\hspace{-0.04truecm}1)\eta
\sinh(2\theta_{p_{n+1}}\hspace{-0.04truecm}-\hspace{-0.04truecm}\eta)}
\overline{K}^+_{11}(m^{(l)},\a^{(l)}|-\theta_{p_{n+1}})\rt\}\no\\
\hspace{-0.4truecm}&\stackrel{(\ref{K+ 11 22 right 1})}{=}&
\hspace{-0.4truecm} 2e^{\theta_{p_{n+1}}}\frac{\sinh
(\hspace{-0.04truecm}-\hspace{-0.04truecm}2\theta_{p_{n+1}}
\hspace{-0.04truecm}+\hspace{-0.04truecm}2\eta)}
{\sinh(\hspace{-0.04truecm}-\hspace{-0.04truecm}2\theta_{p_{n+1}}\hspace{-0.04truecm}+\hspace{-0.04truecm}\eta)}
\sinh(\hspace{-0.04truecm}-\hspace{-0.04truecm}\theta_{p_{n+1}}\hspace{-0.04truecm}-\hspace{-0.04truecm}\a_+)
\cosh(\hspace{-0.04truecm}-\hspace{-0.04truecm}\theta_{p_{n+1}}\hspace{-0.04truecm}-\hspace{-0.04truecm}\b_+)\no\\[4pt]
&&\times F_{n+1}(\theta_{p_1},\cdots,\theta_{p_n},\theta_{p_{n+1}}).\label{Recursive-relation-1}
\eea Iterating the above recursive relation, we arrive at the relations (\ref{Psi-1}).

\subsection*{C.2 The proof of the reference state}

Due to the fact that the particular choice
(\ref{case 1 K-}) of the parameters $m^{(r)},\,\a^{(r)}$ makes the matrix element $K_{21}^-(m^{(r)}+N,\a^{(r)}|u)$ vanishes
(see (\ref{Off-diagonal-K-})), we can derive the following relations from (\ref{C right function}) and (\ref{A right function})
\begin{eqnarray}
\hspace{-0.8truecm}\mathscr{C}_{m^{(r)}}(u|\a^{(r)})|\Omega\rangle\hspace{-0.28truecm}&=&\hspace{-0.28truecm}0,\label{C diagonal right}\\
\hspace{-0.8truecm}\mathscr{A}_{m^{(r)}}(u|\a^{(r)})|\Omega\rangle\hspace{-0.28truecm}&=\hspace{-0.28truecm}&{K^-_{11}}(m^{(r)}+N,\a^{(r)}|u)
\bar A(u)\,|\Omega\rangle.
\label{A diagonal right}
\end{eqnarray}
The definitions (\ref{def U left}) and (\ref{def U+ right}) of the two gauged double-row monodromy matrices
and the relations (\ref{orth M})-(\ref{orth hat M}) allow us to express the operators $\overline{\mathscr{C}}_{m'}(u|\a^{(l)})$ and
$\overline{\mathscr{D}}_{m'}(u|\a^{(l)})$ in terms of some linear combinations of $\mathscr{A}_{m^{(r)}}(u|\a^{(r)})$, $\mathscr{B}_{m^{(r)}}(u|\a^{(r)})$,
$\mathscr{C}_{m^{(r)}}(u|\a^{(r)})$ and $\mathscr{D}_{m^{(r)}}(u|\a^{(r)})$ respectively, namely,
\bea
\hspace{-0.8truecm}\overline{\mathscr{C}}_{m'}(-u|\a^{(l)})\hspace{-0.28truecm}&=&\hspace{-0.28truecm}
      \overline{X}_{m'}(\hspace{-0.04truecm}-\hspace{-0.04truecm}u|\a^{(l)})
      X_{m^{(r)}}(\hspace{-0.04truecm}-\hspace{-0.04truecm}u|\a^{(r)})\mathscr{A}_{m^{(r)}}(-u|\a^{(r)})
      \overline{Y}_{m^{(r)}}(u|\a^{(r)})\widehat{X}_{m'}(u|\a^{(l)}) \no\\
\hspace{-0.8truecm}\hspace{-0.28truecm}&&\hspace{-0.28truecm}+\overline{X}_{m'}(\hspace{-0.04truecm}-\hspace{-0.04truecm}u|\a^{(l)})
      Y_{m^{(r)}\hspace{-0.04truecm}-\hspace{-0.04truecm}2}(\hspace{-0.04truecm}-\hspace{-0.04truecm}u|\a^{(r)})\mathscr{C}_{m^{(r)}}(-u|\a^{(r)})
      \overline{Y}_{m^{(r)}}(u|\a^{(r)})\widehat{X}_{m'}(u|\a^{(l)}) \no\\
\hspace{-0.8truecm}\hspace{-0.28truecm}&&\hspace{-0.28truecm}+\overline{X}_{m'}(\hspace{-0.04truecm}-\hspace{-0.04truecm}u|\a^{(l)})
      X_{m^{(r)}\hspace{-0.04truecm}+\hspace{-0.04truecm}2}(\hspace{-0.04truecm}-\hspace{-0.04truecm}u|\a^{(r)})\mathscr{B}_{m^{(r)}}(-u|\a^{(r)})
      \overline{X}_{m^{(r)}}(u|\a^{(r)})\widehat{X}_{m'}(u|\a^{(l)}) \no\\
\hspace{-0.8truecm}\hspace{-0.28truecm}&&\hspace{-0.28truecm}+\overline{X}_{m'}(\hspace{-0.04truecm}-\hspace{-0.04truecm}u|\a^{(l)})
      Y_{m^{(r)}}(\hspace{-0.04truecm}-\hspace{-0.04truecm}u|\a^{(r)})\mathscr{D}_{m^{(r)}}(-u|\a^{(r)})
      \overline{X}_{m^{(r)}}(u|\a^{(r)})\widehat{X}_{m'}(u|\a^{(l)}),\label{C-expression}\\
\hspace{-0.8truecm}\overline{\mathscr{D}}_{m'}(-u|\a^{(l)})\hspace{-0.28truecm}&=&\hspace{-0.28truecm}
      \overline{X}_{m'}(\hspace{-0.04truecm}-\hspace{-0.04truecm}u|\a^{(l)})
      X_{m^{(r)}}(\hspace{-0.04truecm}-\hspace{-0.04truecm}u|\a^{(r)})\mathscr{A}_{m^{(r)}}(-u|\a^{(r)})
      \overline{Y}_{m^{(r)}}(u|\a^{(r)})\widehat{Y}_{m'+2}(u|\a^{(l)}) \no\\
\hspace{-0.8truecm}\hspace{-0.28truecm}&&\hspace{-0.28truecm}+\overline{X}_{m'}(\hspace{-0.04truecm}-\hspace{-0.04truecm}u|\a^{(l)})
      Y_{m^{(r)}\hspace{-0.04truecm}-\hspace{-0.04truecm}2}(\hspace{-0.04truecm}-\hspace{-0.04truecm}u|\a^{(r)})\mathscr{C}_{m^{(r)}}(-u|\a^{(r)})
      \overline{Y}_{m^{(r)}}(u|\a^{(r)})\widehat{Y}_{m'\hspace{-0.04truecm}+\hspace{-0.04truecm}2}(u|\a^{(l)}) \no\\
\hspace{-0.8truecm}\hspace{-0.28truecm}&&\hspace{-0.28truecm}+\overline{X}_{m'}(\hspace{-0.04truecm}-\hspace{-0.04truecm}u|\a^{(l)})
      X_{m^{(r)}\hspace{-0.04truecm}+\hspace{-0.04truecm}2}(\hspace{-0.04truecm}-\hspace{-0.04truecm}u|\a^{(r)})\mathscr{B}_{m^{(r)}}(-u|\a^{(r)})
      \overline{X}_{m^{(r)}}(u|\a^{(r)})\widehat{Y}_{m'\hspace{-0.04truecm}+\hspace{-0.04truecm}2}(u|\a^{(l)}) \no\\
\hspace{-0.8truecm}\hspace{-0.28truecm}&&\hspace{-0.28truecm}+\overline{X}_{m'}(\hspace{-0.04truecm}-\hspace{-0.04truecm}u|\a^{(l)})
      Y_{m^{(r)}}(\hspace{-0.04truecm}-\hspace{-0.04truecm}u|\a^{(r)})\mathscr{D}_{m^{(r)}}(-u|\a^{(r)})
      \overline{X}_{m^{(r)}}(u|\a^{(r)})\widehat{Y}_{m'\hspace{-0.04truecm}+\hspace{-0.04truecm}2}(u|\a^{(l)}).\label{D-expression}
\eea
The vanishing condition (\ref{Vanishing-cond-1}) implies that
\bea
\langle\a^{(l)},m';\theta_{p_1},\cdots,\theta_{p_n}|\,\overline{\mathscr{C}}_{m'}(-\theta_{p_{n+1}}|\a^{(l)})\,|\Omega\rangle=0,\quad n=0,1,\cdots,N-1.
\eea
Keeping the relations (\ref{C diagonal right}) and (\ref{A diagonal right}) in mind and using the above  equations and the explicit
expressions (\ref{Intertwiner-1}), (\ref{Over-X})-(\ref{Intertwiner-2}), after a tedious calculation, we can derive the following recursive
relations
\bea
{\langle}\a^{(l)},m';\theta_{p_1},\cdots,\theta_{p_{n+1}}|\Omega\rangle&=&K_{11}^-(m^{(r)}+N,\a^{(r)}|-\theta_{p_{n+1}})\bar A(-\theta_{p_{n+1}})\no\\
&&\times {\langle}\a^{(l)},m';\theta_{p_1},\cdots,\theta_{p_{n}}|\Omega\rangle\no\\
&\stackrel{(\ref{Off-diagonal-K-})}{=}&2e^{-\theta_{p_{n+1}}}\sinh(\theta_{p_{n+1}}+\alpha_-)\cosh(\theta_{p_{n+1}}+\beta_-)\bar A(-\theta_{p_{n+1}})\no\\
&&\times {\langle}\a^{(l)},m';\theta_{p_1},\cdots,\theta_{p_{n}}|\Omega\rangle,\no\\
n&=&0,1,\cdots,N-1.
\eea  Iterating the above recursive relations, we have
\bea
{\langle}\a^{(l)},m';\theta_{p_1},\cdots,\theta_{p_{n}}|\Omega\rangle&=&\hspace{-0.4truecm}\prod_{j=1}^n
\lt\{2e^{-\theta_{p_{j}}}\sinh(\theta_{p_{j}}+\alpha_-)\cosh(\theta_{p_{j}}+\beta_-)\bar A(-\theta_{p_{j}})\rt\}
{\langle}\a^{(l)}+m'|\Omega\rangle,\no\\
n&=&0,1,\cdots,N.\no
\eea Comparing the above relations with the conditions (\ref{Reference-state-1}), we conclude that the
state $|\Omega\rangle$  given by (\ref{Reference-state}) is indeed the reference state which we are looking for.
Therefore,  the Bethe state $|\l_1,\cdots,\l_N\rangle$ given by (\ref{Bethe-right}) with the reference state $|\Omega\rangle$ given by
(\ref{Reference-state}) becomes an eigenstate of the transfer matrix $t(u)$ with the
eigenvalue $\L(u)$ given by (\ref{T-Q-1}) provided that the $N$ parameters $\{\l_j|j=1,\cdots,N\}$ satisfy the BAEs (\ref{BAE-1}).

%%%%%%%%%%%%%%%%%%%%%%%%%%%%%%%%%%%%%%%%%%%%%%%%%%%%%%%%%%%%%%%
%                                                             %
%  References                                                 %
%                                                             %
%%%%%%%%%%%%%%%%%%%%%%%%%%%%%%%%%%%%%%%%%%%%%%%%%%%%%%%%%%%%%%%

\end{document}